\documentclass[twocolumn,prb]{revtex4-1}

\usepackage{color}
\usepackage{array}
\usepackage{amsmath}
\usepackage{amssymb}
\usepackage{wasysym}
\usepackage{bm}
\usepackage{graphicx}
\usepackage{txfonts}
\usepackage{epsfig}
\usepackage[latin9]{inputenc}
\providecommand{\tabularnewline}{\\}

\begin{document}

\title{Are there quantum oscillations in an incommensurate charge density
wave?}

\author{Yi Zhang, Akash V. Maharaj, and Steven Kivelson}
\affiliation{Department of Physics, Stanford University, Stanford,
California 94305, USA}
\date{\today}

\begin{abstract}
Because a material with an incommensurate charge density wave (ICDW)
is only quasi-periodic, Bloch's theorem does not apply and there is
no sharply defined Fermi surface. We will show that, as a
consequence, there are no quantum oscillations which are truly
periodic functions of $1/B$ (where $ B$ is the magnitude of an
applied magnetic field). For a weak ICDW, there exist broad ranges
of $1/B$ in which approximately periodic variations occur, but with
frequencies that vary inexorably in an unending cascade with
increasing $1/B$. For a strong ICDW, e.g. in a quasi-crystal, no
quantum oscillations survive at all. Rational and irrational numbers
really are different.
\end{abstract}

\maketitle

\section{Introduction}

Quantum oscillations (QOs) provide powerful experimental avenues to
precisely characterize the Fermi surface structure in an electron
system\cite{Sudip,louis}. At zero temperature and in the limit of
small magnetic field $B$ (where the semiclassical approximation is
asymptotically exact), measurable quantities such as the density of
states (DOS) and various transport properties oscillate as a
function of $1/B$ with a periodicity that is inversely proportional
to the cross-sectional area of the Fermi surface. In the presence of
a commensurate charge density wave (CDW), the QOs can be simply
understood by folding the momentum space Brillouin zone in accord
with the enlargement of the crystal unit cell. In the case of an
incommensurate charge density wave(ICDW), however, the translational
symmetry along the direction of the CDW wave vector (or vectors) is
completely broken; Bloch's theorem no longer applies, and hence the
notion of a Fermi surface is at best an approximate concept.
Nonetheless, it has been widely accepted that new oscillation
periods arise from the reconstructed electron and hole pockets that
result from the perturbative folding of the Fermi surface in a
manner similar to that of the commensurate
case\cite{sebastian2012,sebastian2014,sachdev2014}.

We have extensively studied the problem of QOs in a two-dimensional
system in the presence of a unidirectional ICDW. We obtain numerical
solutions on systems with a linear dimension up to $\sim
O\left(10^6\right)$ sites using an efficient recursive Green's
function method. For weak ICDW potentials $V \ll W$ where $W$ is the
bandwidth, we obtain a perturbative understanding of these results,
while larger values of $V/W$ can be understood by introducing an
exact duality between the two-dimensional system with a
unidirectional ICDW, and a three-dimensional system without the ICDW
but with an additional magnetic flux. For the two-dimensional system
with an ICDW of ordering vector $Q$ and amplitude $V$, the dual
three-dimensional system has a hopping matrix element $V/2$ in the
out-of-plane ($\hat{z}$) direction, and a ``tilted'' magnetic field
$ \vec{B}^{eff}=B\hat{z} + (Q/2\pi)\hat{y}$ where $\hat{x}$ and
$\hat{y}$ are the in-plane directions parallel and perpendicular to
the charge ordering vector (where magnetic fields are measured in
units of a flux quanta per unit cell). This mapping is readily
generalized to other dimensions, as well as to the case of
bidirectional ICDW order.

We find that there is no limit in which perfectly periodic QOs occur
in the presence of an ICDW. In the strong coupling limit, for $V/W$
greater than a critical value of order $1$, there are no well
defined QOs at all. In the dual system, the critical $V/W$ is
identified with the point at which the underlying three-dimensional
Fermi surface transforms from a quasi-two-dimensional cylinder to a
closed three-dimensional surface. This limit likely applies to the
case of quasi-crystals in which the incommensurate potential is not
small in any sense. By contrast, in the small $V/W$ limit,
oscillations that approximate the QOs of a true crystal are observed
over large but finite intervals of $1/B$, but the observed period
varies depending on the range of $B$ studied.

The nature of the QOs for small $V/W$, and the manner in which they
deviate from strict periodicity can be understood intuitively in
terms of the conventional theory of magnetic breakdown as a
perturbative effect. Specifically, there is a hierarchy of gaps on
the Fermi surface, $\Delta_n \sim W(V/W)^n$, whose magnitudes are
governed by the lowest order in perturbation theory at which they
appear. Depending on the position of the unperturbed Fermi surface
and its relation to the CDW ordering vector, only some of these gaps
open on the Fermi surface, and so it is only these that matter. The
hierarchy of gaps on the Fermi surface terminates at $n_{\rm
max}\leq q$ for a $q^{th}$ order commensurate CDW, but it continues
to arbitrarily high $n$ in the incommensurate case. In a given range
of $B$, magnetic breakdown effectively eliminates the effect on the
electron dynamics of all gaps that are sufficiently small that
$\Delta^2 /W \ll \omega_c $, where $\omega_c \sim W B$ is the
effective cyclotron energy, while all gaps that are large compared
to $\sqrt{\omega_cW}$ are respected by the semiclassical dynamics.
If $n=N$ and $n=N+M$ are two sequential terms in this hierarchy,
then for small $V/W$ there exists a parametrically broad range of
fields, $(\Delta_N/W)^2 \gg B \gg (\Delta_{N+M}/W)^2$, in which all
gaps of order $n \geq N+M$ can be neglected, but the Fermi surface
is effectively reconstructed by all gaps with $n \leq N$. However,
inevitably, a further reconstruction of the Fermi surface must occur
in some order of perturbation theory, giving a fractal character to
the QO spectrum.

The rest of this paper is organized as follows. Sec.
\ref{sec:CDW_infield} reviews the canonical understanding of QOs in
the presence and absence of CDWs. In Sec. \ref{sec:model} we
introduce the tight binding model which forms the basis of our
numerical analysis, and also discuss its duality properties.
Numerical results are presented in Sec. \ref{sec:results}, where we
show the field dependence of QO periods for a commensurate CDW and a
small amplitude ICDW, along with the spectacular absence of QOs for
a large amplitude ICDW. We also explain the eventual breakdown of
the perturbative picture for the ICDW QOs. In Sec. \ref{dual} we
numerically verify the exactness of the duality picture, using this
view to uncover several novel properties of ICDW in magnetic fields.
Finally, Sec. \ref{sec:generalization} discusses further
applications of this model to localization problems in one dimension
as well as the case of bidirectional ICDW order.

\section{Charge density waves in a magnetic field}\label{sec:CDW_infield}

\subsection{Quantum oscillations in ordinary metals}

Quantum oscillatory phenomena can be generally understood to arise
from the semi-classical quantization of electronic energies in an
applied magnetic field -- which is asymptotically exact in the $B
\to 0$ limit. The dynamics of Bloch electrons in an applied magnetic
field is given semi-classically by the Lorentz force law:
\begin{equation}
\hbar \frac{d\vec{k}}{dt}=-e\left[\vec{v}(\vec{k})\times\vec{B}\right]=\frac{e}{\hbar}\left(\vec{B}\times \frac{dE(\vec{k})}{d\vec{k}}\right),
\end{equation}
where $\vec{v}(\vec{k})=(1/\hbar)\,dE/d\vec{k}$ is the electron
group velocity. This implies $\vec{k}\cdot\vec{B}=0$ and
$dE(\vec{k})/dt=0$, i.e. the electrons move in orbits of constant
energy in planes perpendicular to the magnetic field.

In the absence of any Berry phase, the quantum phase the electrons
accrue on each orbit around the Fermi surface is $\hbar S_{k}/eB$,
where $S_{k}$ is the area of Fermi surface cross section
perpendicular to the applied magnetic field. Thus the maximum and
minimum values of $S_{k}$ govern the interference between multiple
trajectories. Maxima in the semiclassical DOS (corresponding to the
point at which a Landau level just crosses the Fermi energy) occur
whenever
\begin{equation}
\frac{1}{B}=\left(n+\frac{1}{2}\right)\frac{4\pi^{2}}{\Phi_{0}S_{k}}
\label{eq:rescondi}
\end{equation}
for integer $n\in\mathbb{Z}$ where $\Phi_{0}=h/e$ is the magnetic
flux quantum. For fixed Fermi energy (and hence constant $S_k$) this
leads to perfectly periodic QOs with period $\Delta(1/B)=2\pi
e/\hbar S_{k}$.

\subsection{Commensurate CDWs and magnetic breakdown}\label{sec:breakdowns}

\begin{figure}
\begin{centering}
\includegraphics[scale=0.4]{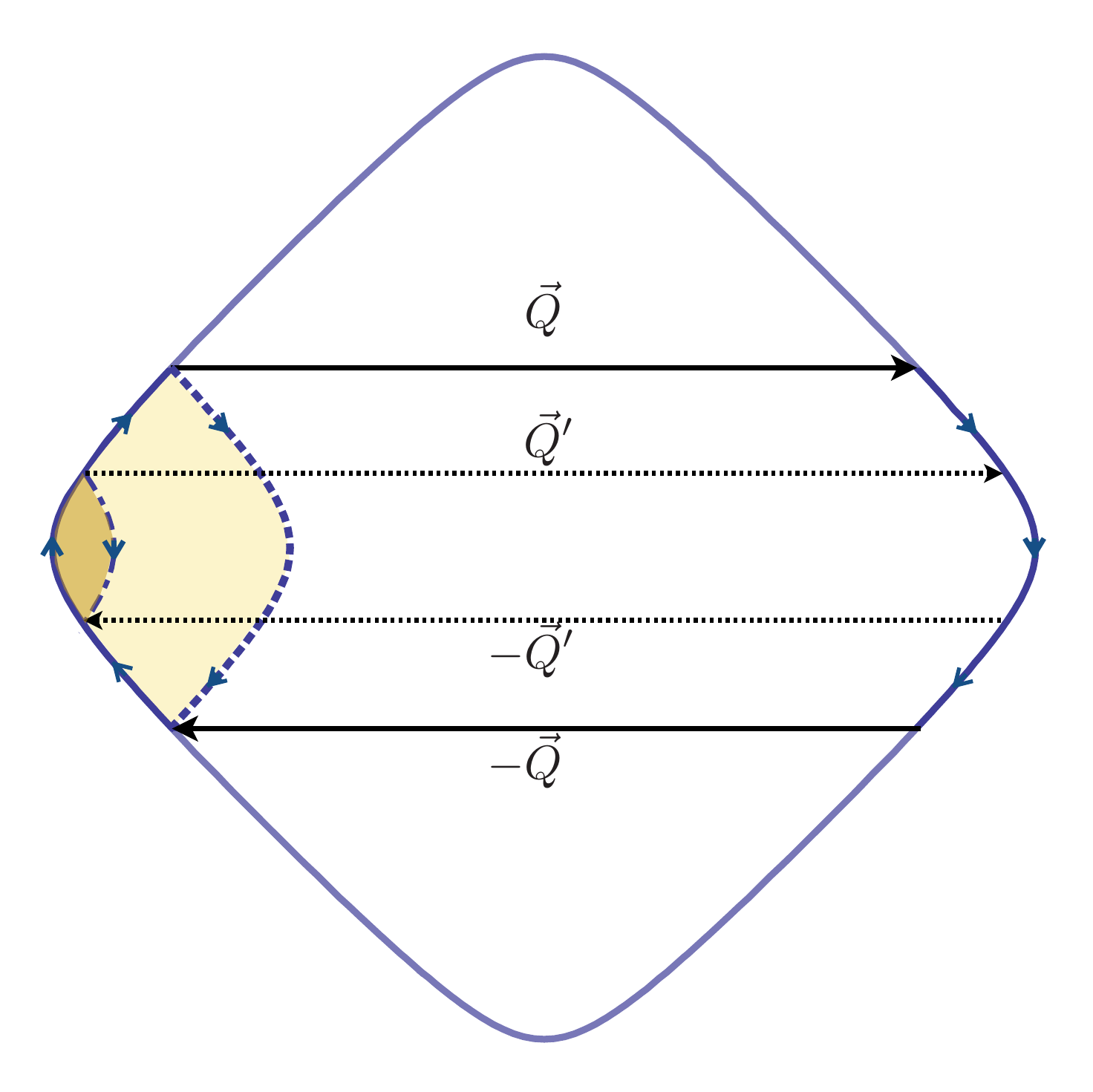}
\protect\caption{The Fermi surface reconstruction with a momentum
transfer of $\pm Q$ by scattering from a CDW is shown as the
yellow-colored region (both the dark and light yellow regions). In
addition, a higher order scattering with a momentum transfer of $\pm
Q'$ may lead to a smaller pocket (dark yellow region) and so on.
Consequently, there may be a hierarchy of Fermi surface pockets
protected by gaps of different sizes.} \label{fig2}
\end{centering}
\end{figure}

The above arguments can be straightforwardly applied to systems in
which the translation symmetry of the crystal is spontaneously
broken by a commensurate CDW with wave-vector $\vec Q= Q\hat x$
where, in units in which the lattice constant is $a=1$, $Q=2\pi p/q$
with $p,q\in\mathbb{Z}$ relatively prime. This simply defines a new
crystal structure with $q$ times as large a unit cell and $q$ times
as many bands in a folded Brillouin zone which is $q$ times smaller
in the $\hat x$ direction. Generically,\cite{norman,yao,chakravarty}
the CDW causes a ``reconstruction'' of the Fermi surface due to gaps
that open at the new Brillouin zone-boundary, resulting in smaller
electron and hole pockets (plus open Fermi surfaces that do not
contribute to the QOs), thus giving rise to new periodicities in the
QOs. The reconstruction can be viewed as arising from processes in
which the electrons on the Fermi surface are scattered by the CDW
with momentum transfer $\pm nQ\hat x$ modulo $2\pi\hat x$. In the
case of a weak sinusoidal density wave, $V/W \ll 1$, there is a
hierarchy of scales associated with these processes depending on
$n$, as such processes arise first in $n^{th}$ order perturbation
theory, see Fig. \ref{fig2}.

In the notation used in the introduction, gaps indexed in this way
depend parametrically on $V$ as $\Delta_n = \alpha W (V/W)^n\left[ 1
+ {\cal O}(V/W)^2\right]$ where $\alpha$ does not depend on $V$.
Depending on the relation between the Fermi surface location and the
ordering vector $Q$, gaps with a given index $n$ may not involve
states at the Fermi surface, in which case they can be neglected for
present purposes. For instance, if $Q^\prime$ in Fig. \ref{fig2}
were slightly longer, it would fail to span the indicated Fermi
surface, and so would not produce any Fermi surface reconstruction.
Notice that the area of the reconstructed Fermi surface pockets is
purely geometric in the limit of vanishingly small $V/W$, determined
solely by the structure of the underlying Fermi surface and the
specified CDW ordering vector, but for finite $V$ the opening of
gaps causes a non-zero displacement of the Fermi surface which can
produce changes in the area of the various pockets of order
$\Delta_n/W$. This effect becomes qualitatively significant if $V/W$
is not small, in which case the perturbative approach must be
abandoned.

For finite field strength $B$ corrections to semiclassical
quantization arise due to `magnetic breakdown', associated with
transitions between semiclassical trajectories. It is easy to
show\cite{cohen1961,blount1960,kee2010} that magnetic breakdown is
negligible so long as $\ell\delta k \gg 1$, where $\ell =
\sqrt{\Phi_0/2\pi B}$ is the magnetic length and $\delta k$ is the
distance of closest approach in reciprocal space between two
semi-classical trajectories. (The precise criterion depends on the
local curvature of the trajectories as
well\cite{chambers1966,kee2010}.) Alternatively, for weak CDW order,
this criterion can be expressed as $\omega_c \ll \Delta^2/W$, where
$\omega_c$ is the cyclotron energy and $\Delta$ is the relevant gap
induced by the CDW. When there exists a series of Fermi surface
pockets produced by gaps of different sizes from multiple orders of
reconstructions, this results in a corresponding hierarchy of
breakdown magnetic fields.

\subsection{Incommensurate CDW in a magnetic field}

When the period of charge modulation is not a rational multiple of
the underlying lattice spacing, {\it i.e.} when $Q = 2\pi \alpha$
where $\alpha$ is an irrational number, the CDW is referred to as
incommensurate. While many properties of ICDW materials can be
understood by treating the CDW potential perturbatively, or better
still by approximating $\alpha$ by a nearby commensurate
approximant, $p/q\approx \alpha$, we will show in this paper that an
exact treatment of the problem reveals a number of fundamental
properties that are not captured by this type of approximation.

\section{The Model}\label{sec:model}

To begin with, we define a general model on a $d$-dimensional
hyper-cubic lattice in the presence of a uniform magnetic field and
a CDW potential:
\begin{eqnarray}
\hskip-10pt H=-\sum_{\langle \vec r,\vec{r^\prime}\rangle}\left[
t_{\vec r-\vec r^\prime}\exp\left[i A(\vec r,\vec r^\prime)\right]
c_{\vec{r}}^{\dagger}c^{}_{\vec{r}^\prime}+ {\rm h.c.}\right]
-\sum_{\vec r}U(\vec r)c_{\vec{r}}^{\dagger}c^{}_{\vec{r}} \label{H}
\end{eqnarray}
where $c_{\vec r}^\dagger$ creates an electron on site $\vec r$,
$\langle\vec r,\vec{r^\prime}\rangle$ designates nearest-neighbor
sites, $t_\nu\equiv t(\pm{\hat e_\nu})$ is assumed real and
positive, but can depend on direction $\hat e_{\nu}$ with $\nu$
running from $1$ to $d$ (with the convention $\hat e_1=\hat x$,
$\hat e_2=\hat y$, etc.), the magnetic flux through any plaquette in
units in which the flux quantum is $\Phi_0=2\pi$ is given by the sum
of $A$ around the plaquette, and by definition $A(\vec r,\vec
r^\prime)=-A(\vec r^\prime,\vec r)$. In all cases, we will assume
the flux is spatially uniform and penetrates only plaquettes that
are parallel to $\hat x$.  We are thus free to chose a gauge that
preserves translational symmetry in all but the $\hat x$ direction,
$A(\vec r,\vec r+\hat e_\nu)=\Phi_\nu\  \vec r \cdot \hat x =
\Phi_\nu\ x$, where, for example, $\Phi_2\equiv \Phi_y=\Phi$ is the
flux through each plaquette in the $x-y$ plane, and $\Phi_1\equiv
\Phi_x=0$.  For the present, we will also consider a unidirectional
CDW in the $\hat x$ direction which consists of a small number $m$
of distinct Fourier components,
\begin{equation}
U(\vec r) = \sum_{j=1}^m V_j\cos[Q_j x - \theta_j].
\end{equation}
Here $\theta_j$ is the relative phase between a given component of
the CDW and the underlying lattice, and $x_j\equiv \theta_j/Q_j$ can
be interpreted as the location of the minimum of the corresponding
CDW potential in Eq. \ref{H} in continuum space.

Because the CDW is unidirectional and we have chosen the appropriate
gauge, we can exploit translation invariance to Bloch diagonalize
the Hamiltonian by Fourier transform perpendicular to $\hat x$. We
define
\begin{equation}
a_{x,\vec k} \equiv (N_\perp)^{-d/2}\sum_{\vec r_\perp}e^{i\vec
k\cdot \vec r_\perp} c_{\vec r_\perp+x\hat x}
\end{equation}
where $\vec r_{\perp}\equiv \vec r - x \hat x$ and $\vec k$ is the
corresponding $d-1$ dimensional Bloch vector. In terms of these,
$H=\sum_{\vec k} H_{\vec k}$ with
\begin{eqnarray}
\label{1d}
H_{\vec k} =&-& \sum_x t_x\left( a_{x,\vec k}^\dagger a_{x+1,\vec k} +{\rm h.c.}\right) \\
&+& \left[2\sum_{\nu=2}^d t_\nu\cos(\Phi_\nu x - k_\nu)
+\sum_{j=1}^m V_j\cos(Q_jx -\theta_j)\right] a_{x,\vec k}^\dagger
a_{x,\vec k} \nonumber
\end{eqnarray}

Thus, for given $\vec k$, the generic problem is equivalent to a
problem in one dimension. Moreover, it is immediately apparent from
Eq. \ref{1d} that there is a formal equivalence between higher
dimensional problems with a magnetic field and lower dimensional
problems with more components of the CDW; e.g. with all $V_j = 0$
there is the familiar relationship between the Hofstadter problem of
a two-dimensional crystal in the presence of a uniform flux $\Phi$
per plaquette and the corresponding one-dimensional problem
(governed by Harper's or the ``almost Mathieu''
equation\cite{aubry1979,sokoloff1985}) of a particle in a sinusoidal
potential.\footnote{Harper's equation is a specific realization of
the almost Mathieu equation, with $t_{y} = t_{x}$.}

\begin{figure}
\begin{centering}
\includegraphics[scale=1.0]{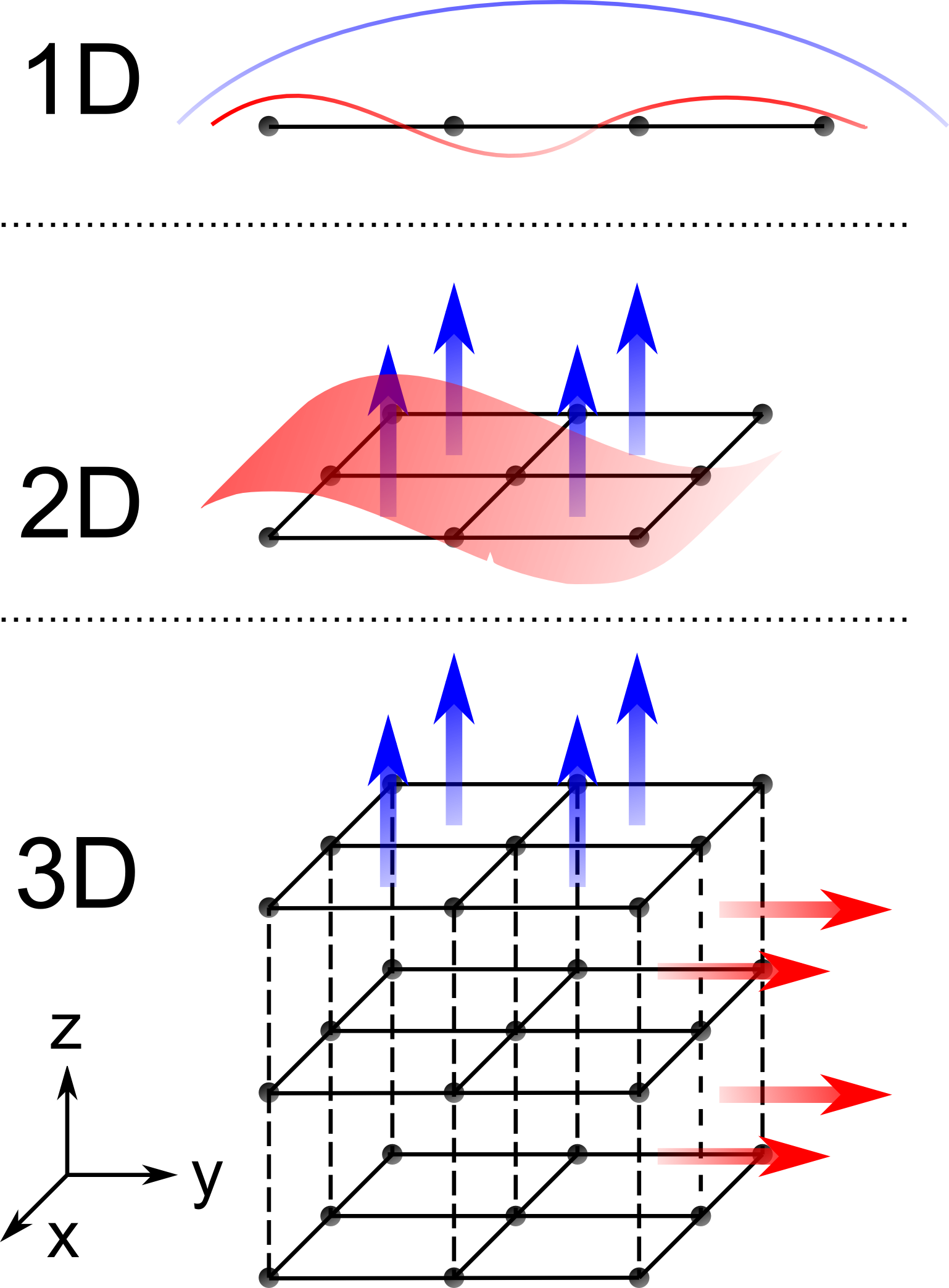}
\protect\caption{Upper panel: A one-dimensional model with two
incommensurately periodic potentials, one with amplitude $2$ and
wave vector $\Phi$ (blue curve) and the other with amplitude $V$ and
wave vector $Q$ (red curve). Middle panel: The equivalent
two-dimensional square lattice model with an ICDW (red curve) and a
magnetic flux (blue arrow) $\Phi=2\pi B$ per plaquette. Lower panel:
The effective cubic lattice model in three dimensions. Here the
hopping amplitudes are 1 along the $\hat x$ and $\hat y$ directions
(solid lines), and $V/2$ along the $\hat z$ direction (dashed
lines), and an additional magnetic flux of $Q$ is now present
through the $xz$ plaquettes (the red arrow). } \label{fig8}
\end{centering}
\end{figure}

{\bf Duality relations:} The problem of present interest is that of
the two-dimensional crystal subjected to a uniform flux $\Phi$ and a
unidirectional CDW potential $V\cos[Qx-\theta]$.  We now see that
this is equivalent both to a one-dimensional system with the doubly
periodic potential, $U(\vec r)=V\cos[Qx-\theta]+2t_y\cos[\Phi x -
k_y]$, and the anisotropic three-dimensional crystal with hopping
matrix elements $t_x$, $t_y$ and $t_z=V/2$ in, respectively, the
$\hat x$, $\hat y$, and $\hat z$ directions subjected to a uniform
magnetic field with flux $\Phi$ through each $xy$ plaquette and flux
$Q$ through each $xz$ plaquette (see Fig. \ref{fig8}). We will use
the one-dimensional representation of the problem as the basis of
our numerical study. However, we gain physical insight into the
solution and approximate analytic understanding by viewing it as a
translationally invariant (crystalline) three-dimensional problem in
a tilted magnetic field. We also note that higher harmonic
components of the CDW potential with wave vectors $nQ$, $n\in
\mathbb Z$, correspond to the same net magnetic field but further
neighbor hoppings in the $\hat z$ direction thus more generic $k_z$
dispersions.

There is still the issue of the sum over $\vec k$. However, for the
one-dimensional problem with a multi-component ICDW potential for
which $Q_i/Q_j$ is irrational for all $i\neq j$, it is easy to see
and straightforward to prove that {\it in the thermodynamic limit},
the spectrum is independent of the the phases $\theta_j$, that is to
say that there is a sliding symmetry for each component of the CDW.
 Translated to the magnetic flux problem, since $\Phi_\nu$
is generically an irrational multiple of $\Phi_0$, this implies that
(except in fine tuned circumstances), the spectrum of Eq. \ref{1d}
will be independent of $\vec k$.  Thus, for all thermodynamic
quantities, the sum over $\vec k$ simply produces a degeneracy
factor of $N_\perp$ for each eigenstate.

{\bf Notation and Units:} We will henceforth suppress the index
$\vec k$ on all quantities, ({\it e.g.} $a_{x,\vec k}$ will be
represented as $a_x$). We will specialize to the case in which the
two-dimensional band structure is isotropic, $t_x=t_y=t$, and will
use units of energy such that $t=1$. Recall that we have adopted
units of length such that the lattice constant is $1$, and units of
magnetic field $B=\Phi/2\pi$ such that $\Phi_0=2\pi$.

\section{Results of Numerical Experiments}\label{sec:results}
In our numerical studies, we compute two physical quantities which
are related %
to the Green's function for the doubly incommensurate
one-dimensional problem with open boundary conditions, Eq. \ref{1d}
with $d=2$ and $n=1$. The Green's function for each $k_y$ is defined
as the matrix inverse:
\begin{align}
G_{k_y}(x,x^{\prime}) = \left(\mu + i \delta -
H_{k_y}\right)^{-1}_{x,x^{\prime}}
\end{align}
where $\mu$ is the chemical potential and $i\delta$ is a very small
imaginary part to round off the singularities, which for large
systems can always be chosen so that, to any desired accuracy, the
results are $\delta$ independent. We calculate both the DOS $\rho$
at the Fermi level
\begin{align}
\rho_{k_y}(\mu) = -\frac{1}{\pi L} \sum_{x} \text{Im}G_{k_y}(x,x)
\end{align}
and the localization length $\lambda$ that defines the exponential
decay of the Green's function $G_{k_y}(1,L) \sim
\exp(-L/\lambda_{k_y})$ between two ends of the system. Further
details on our numerical methods are presented in the Appendix
\ref{sec:numerics}.

\subsection{Results for a commensurate CDW}

We first benchmark our methods on the same model with no CDW ($V=0$)
or a commensurate CDW ($Q=\pi$, $V=0.16$). We set the chemical
potential $\mu=-0.2$, corresponding to slight hole doping away from
half filling to avoid complications that arise from (fine-tuned)
Fermi surface nesting on a bipartite lattice when $\mu=0$; our
conclusions are readily generalizable to other chemical potentials
and shapes of the Fermi surface.

\begin{figure}
\begin{centering}
\includegraphics[scale=0.3]{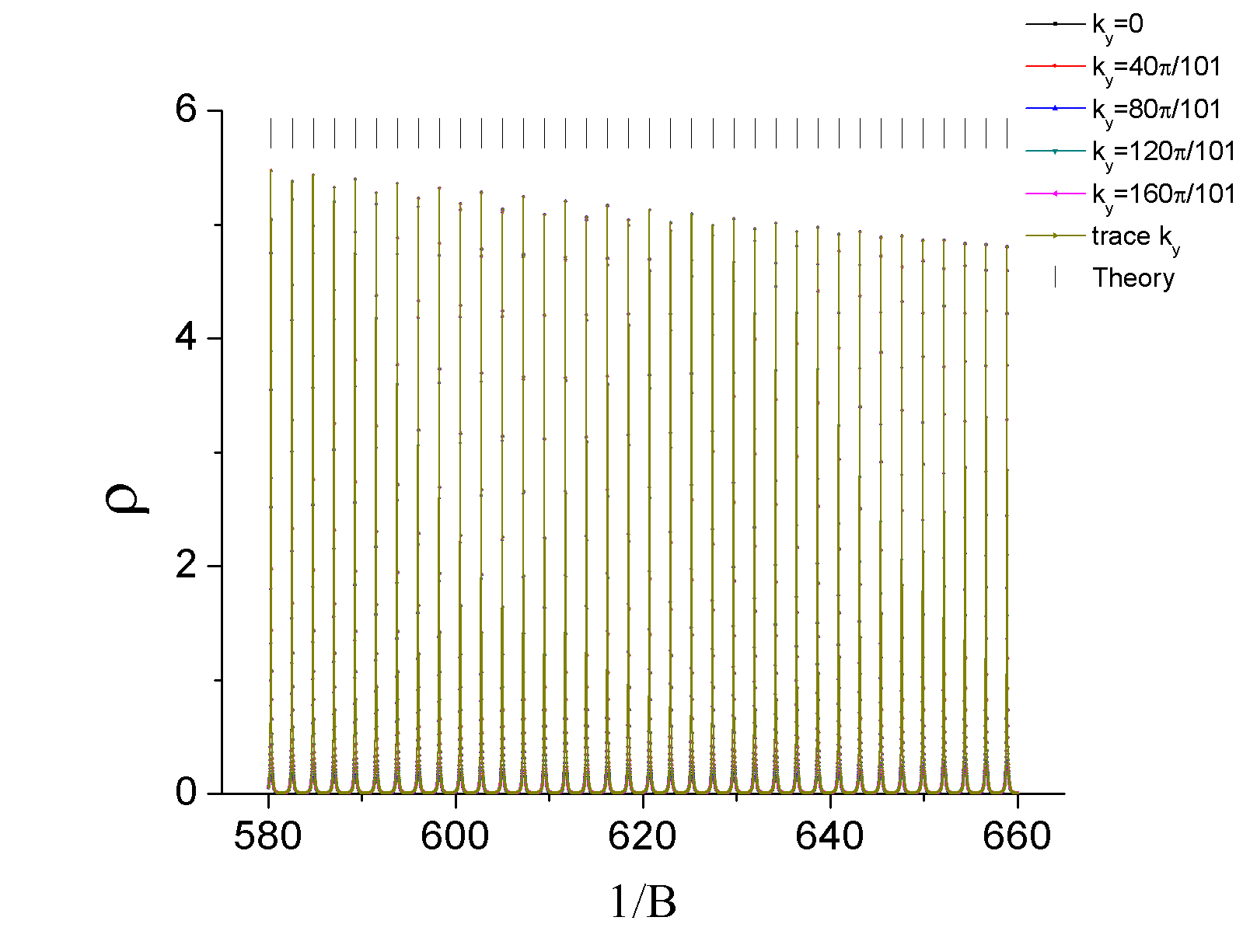} \includegraphics[scale=0.3]{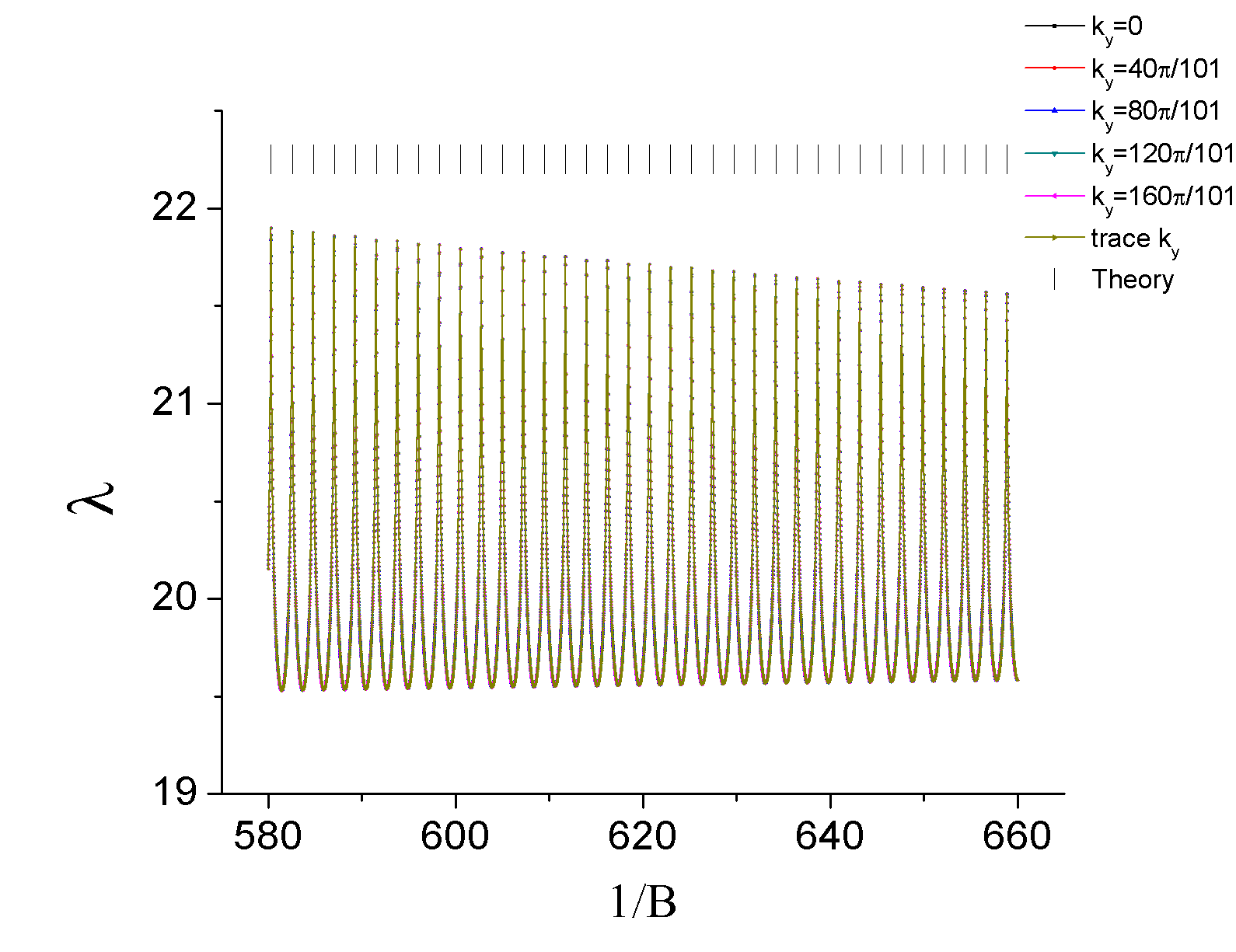}
\caption{(Upper panel) the DOS $\rho$ and (lower panel) the
localization length $\lambda$ versus $1/B$ calculated from various
values of $k_{y}$ as well by as tracing/averaging over all $k_{y}$
(taking $L_y=101$) for a system of length $L=10^5$ without a CDW
$(V=0)$. For comparison, the marks at the top of each figure are the
theoretical values of interference maxima based on the Fermi surface
area of $S_k=0.44546S_{BZ}$.} \label{fig1}
\end{centering}
\end{figure}

To study the QOs for $V=0$, we extract the localization length
$\lambda$ and DOS $\rho$ over a range of magnetic field on a
$L=10^{5}$ system. As shown in Fig. \ref{fig1}, both the DOS $\rho$
(upper panel) and the localization length $\lambda$ (lower panel)
exhibit a clear single-period oscillation. By linearly fitting the
locations of the peaks we obtain the period
$\Delta\left(1/B\right)=2.2452$, which corresponds to a momentum
space area:
\begin{align}
S_{k} = \frac{4\pi^2}{\Delta\left(1/B\right)}=0.4454S_{BZ}
\end{align}
where $S_{BZ}=\left(2\pi\right)^{2}$ is the area of the entire
Brillouin zone of the square lattice. For comparison, given the
zero-field dispersion relation of Eq. \ref{H} with the assumed value
of the chemical potential, the Fermi surface enclosed area in the
absence of any CDW is $S_{k}=0.44546S_{BZ}$. To highlight the high
accuracy of our results, we have also marked the theoretically
expected locations of the maxima from Eq. \ref{eq:rescondi} at the
top of each figure.

We have checked the expected $k_y$ independence of the results by
repeating the calculation for different choices of $k_y$ as well as
by averaging over $k_y$ as shown in Fig. \ref{fig1}. Henceforth we
suppress the $k_y$ label for both $\rho$ and $\lambda$.

\begin{figure}
\begin{centering}
\includegraphics[scale=0.15]{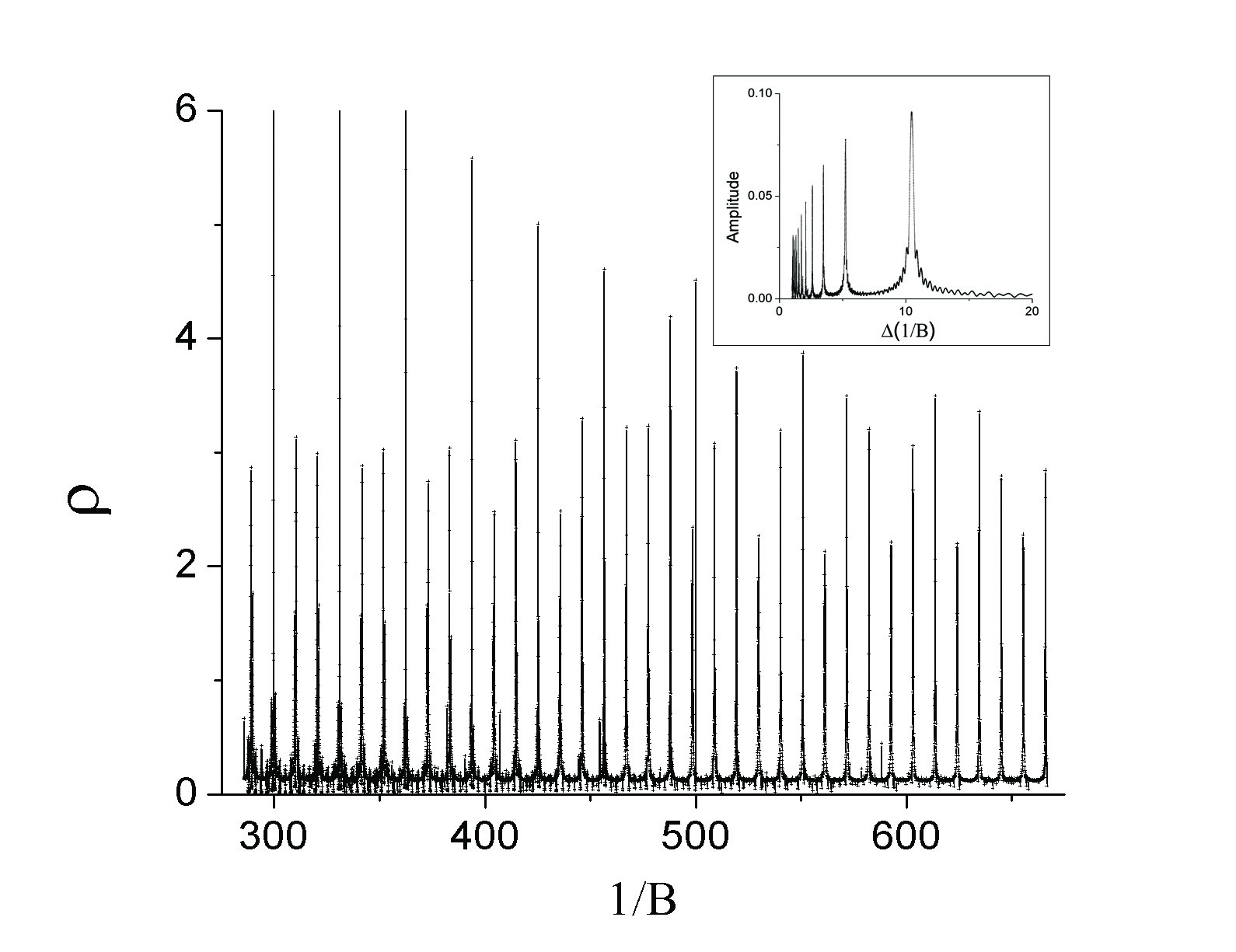}
\includegraphics[scale=0.15]{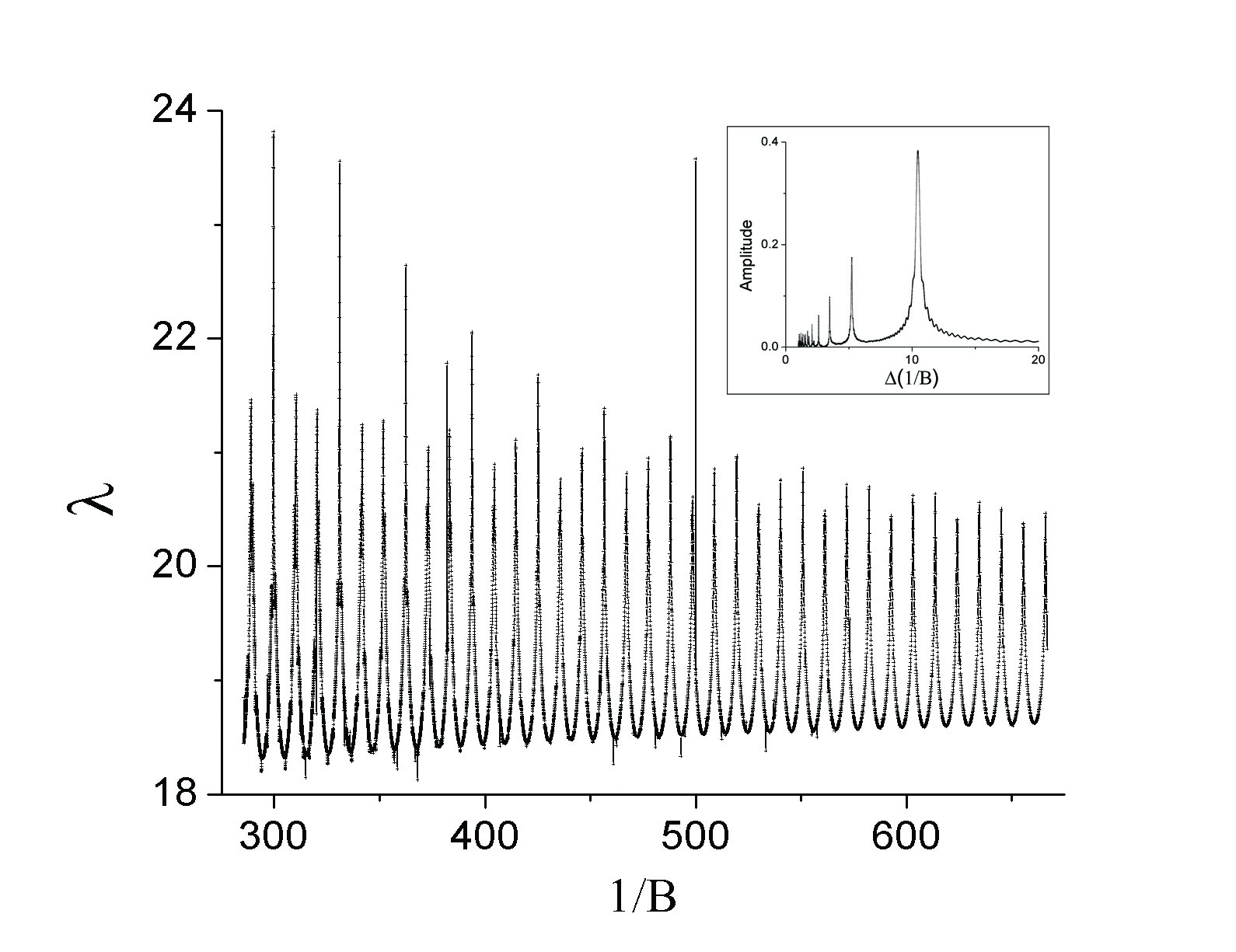}
\includegraphics[scale=0.3]{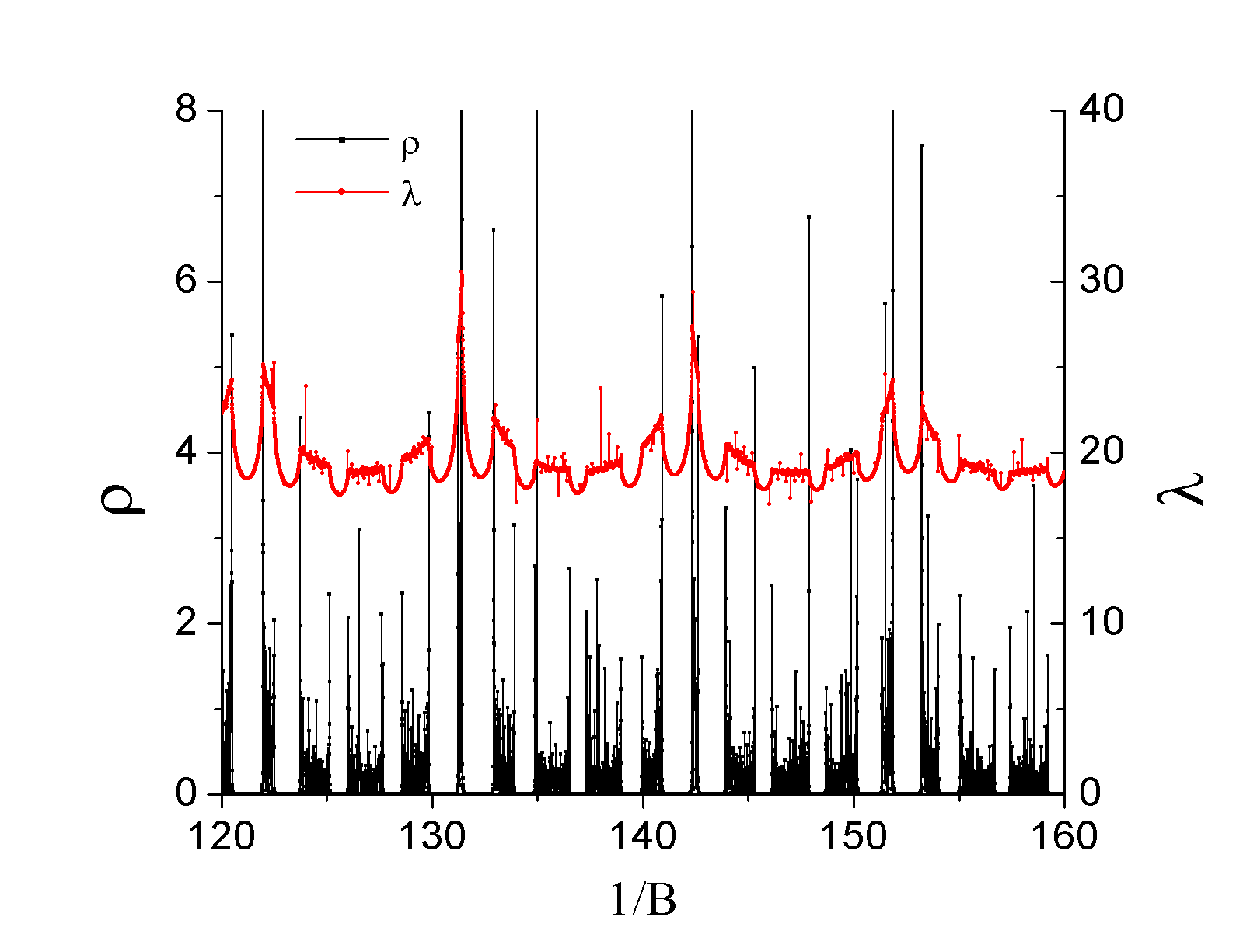}
\protect\caption{The DOS $\rho$ (upper panel) and the localization
length (middle panel) versus the inverse magnetic field $1/B$ for a
system of length $L=2\times10^{5}$ with parameters $\mu=-0.2$,
$V=0.16$ and $Q=\pi$. Insets: the Fourier transform of the QOs
exhibits a clear peak at period $\Delta(1/B)=10.464$ corresponding
to the fundamental period. Lower panel: the QOs behavior of the DOS
and the localization length over a range of larger magnetic fields
$B$.} \label{fig3}
\end{centering}
\end{figure}

As another benchmark, we have studied QOs in the presence of a
commensurate CDW: $V=0.16$ and $Q=\pi$ for $L=2\times10^{5}$; the
results for the DOS and the localization length are shown in the
upper and middle panels of Fig. \ref{fig3}, respectively. A Fourier
transform (Fig. \ref{fig3} insets) reveals peaks and higher
harmonics\footnote{The origin of the higher harmonics can be
attributed to non-linear dependence of the physical quantities on
the orbital phase factor.} that correspond to a fundamental period
of $\Delta\left(1/B\right)=10.464$. This, in turn, reflects a
momentum space area $S_k=0.1911S_{BZ'}$, where $S_{BZ'}=S_{BZ}/2$ is
the area of the folded Brillouin zone according to the enlarged unit
cell. For the given parameters, the Fermi surface consists of only
one closed pocket with an enclosed area of $S_{k}=0.1909S_{BZ'}$.

Strictly speaking, at any finite magnetic field, the QO periods are
magnetic-field dependent due to corrections to the semiclassical
results associated with magnetic breakdown. This behavior is
illustrated in the lower panel of Fig. \ref{fig3}, which is the same
model but with larger values of $B$ ($1/B\in[120, 160]$). QOs with a
period $\Delta\left(1/B\right)=2.245$ are clearly seen, which
corresponds to the Fermi surface area in the absence of the CDW
potential ($V=0$). While oscillations with a period corresponding to
the reconstructed Fermi surface, $\Delta\left(1/B\right)=10.463$,
are also vaguely discernible in this field range; they become
dominant only at smaller fields.

\subsection{Results for an ICDW}
We now consider systems with ICDW, and discuss the consequences
within various regimes of magnetic field strength and CDW amplitude.

\subsubsection{Reconstructed Fermi-surface in the perturbative regime}

Many studies\cite{sebastian2012,sebastian2014,sachdev2014} of QOs in an ICDW use perturbative
arguments to evaluate the reconstructed Fermi surfaces. As indicated
before, this approach is partially justified in the limit of weak
CDW potentials $V \ll t$, where the result can be visualized
dynamically as electrons precessing in semiclassical orbits around a
closed Fermi surface, and only occasionally picking up momentum $\pm
Q$ to jump from one section of the Fermi surface to another.

\begin{figure}
\begin{centering}
\includegraphics[scale=0.155]{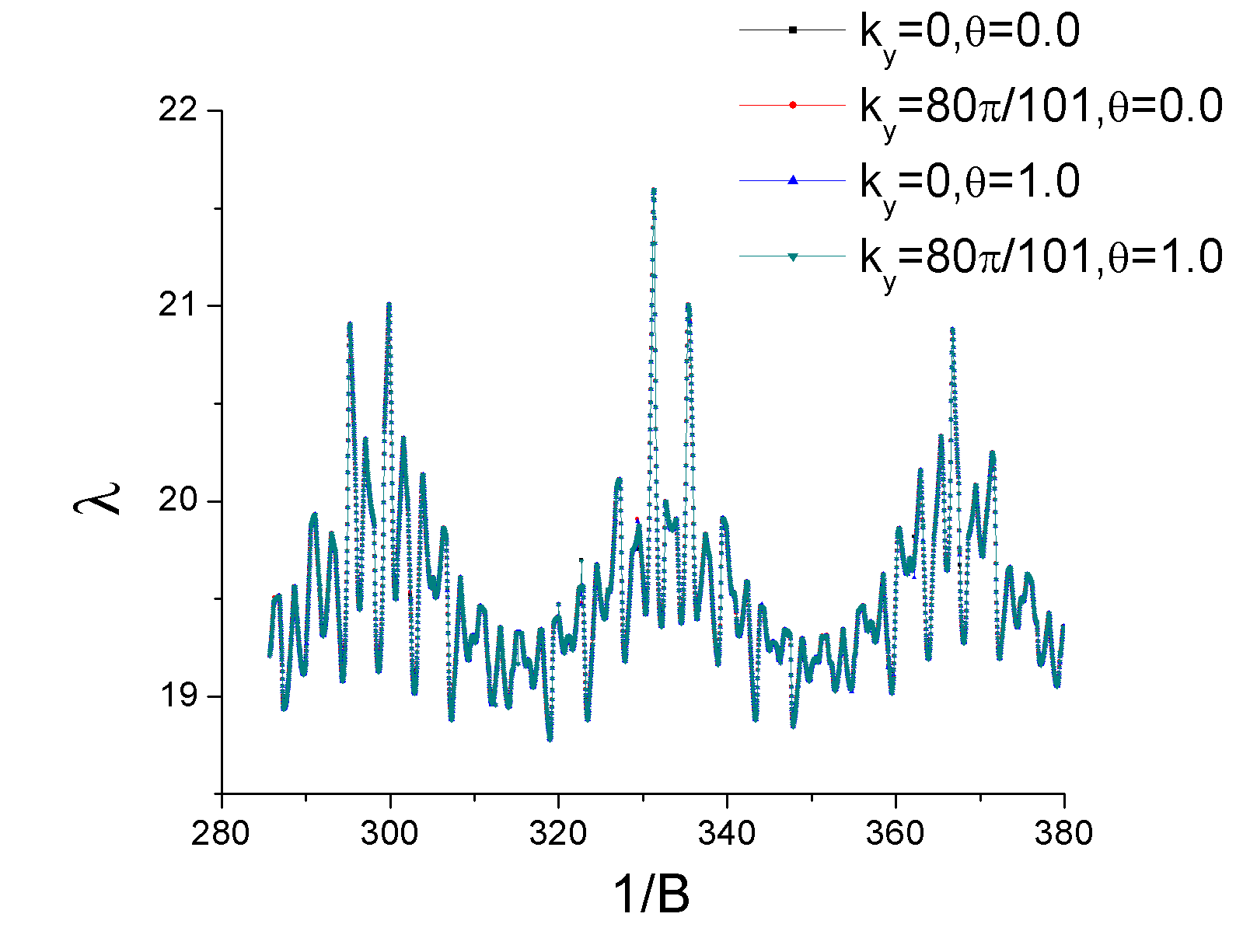}
\includegraphics[scale=0.155]{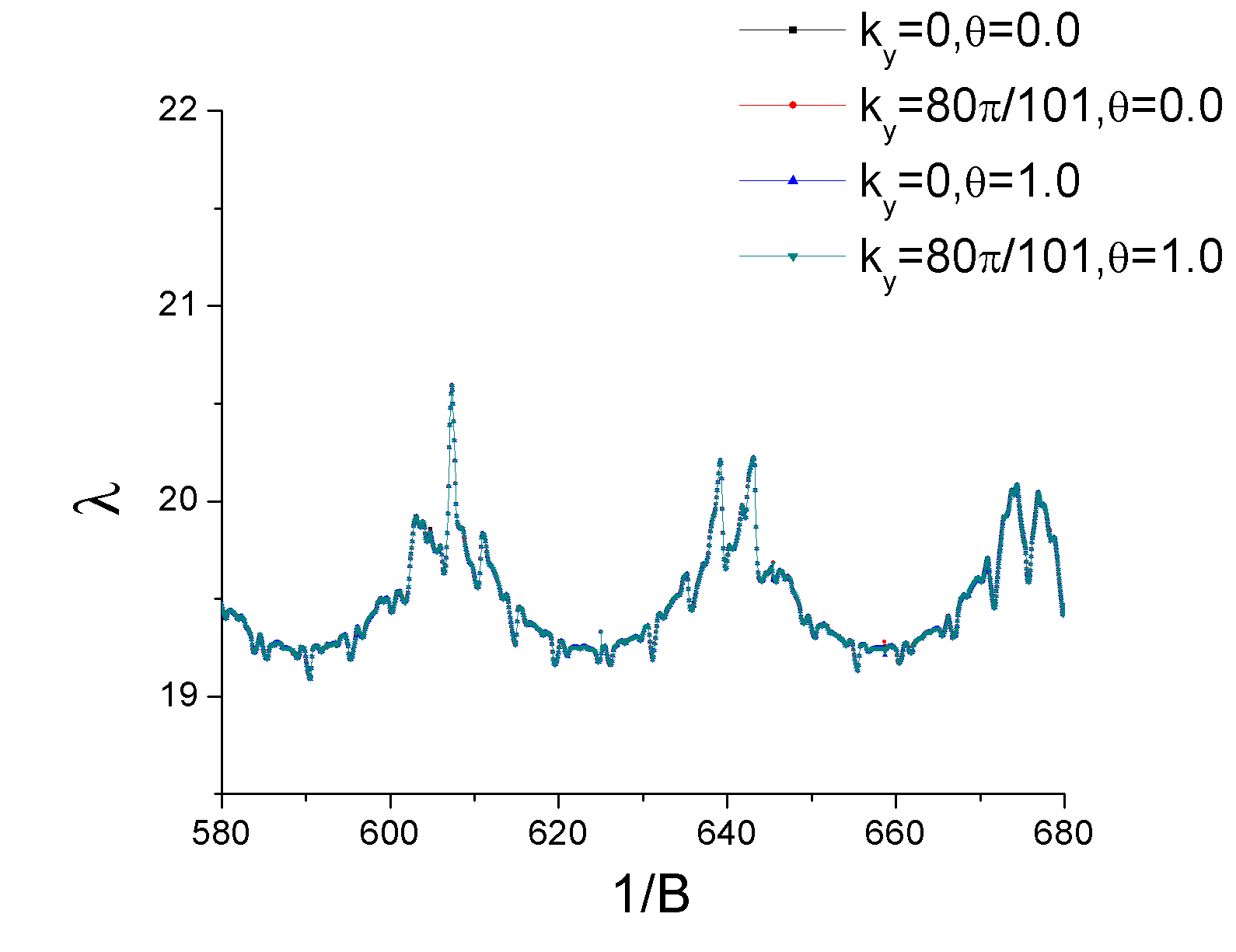}
\includegraphics[scale=0.155]{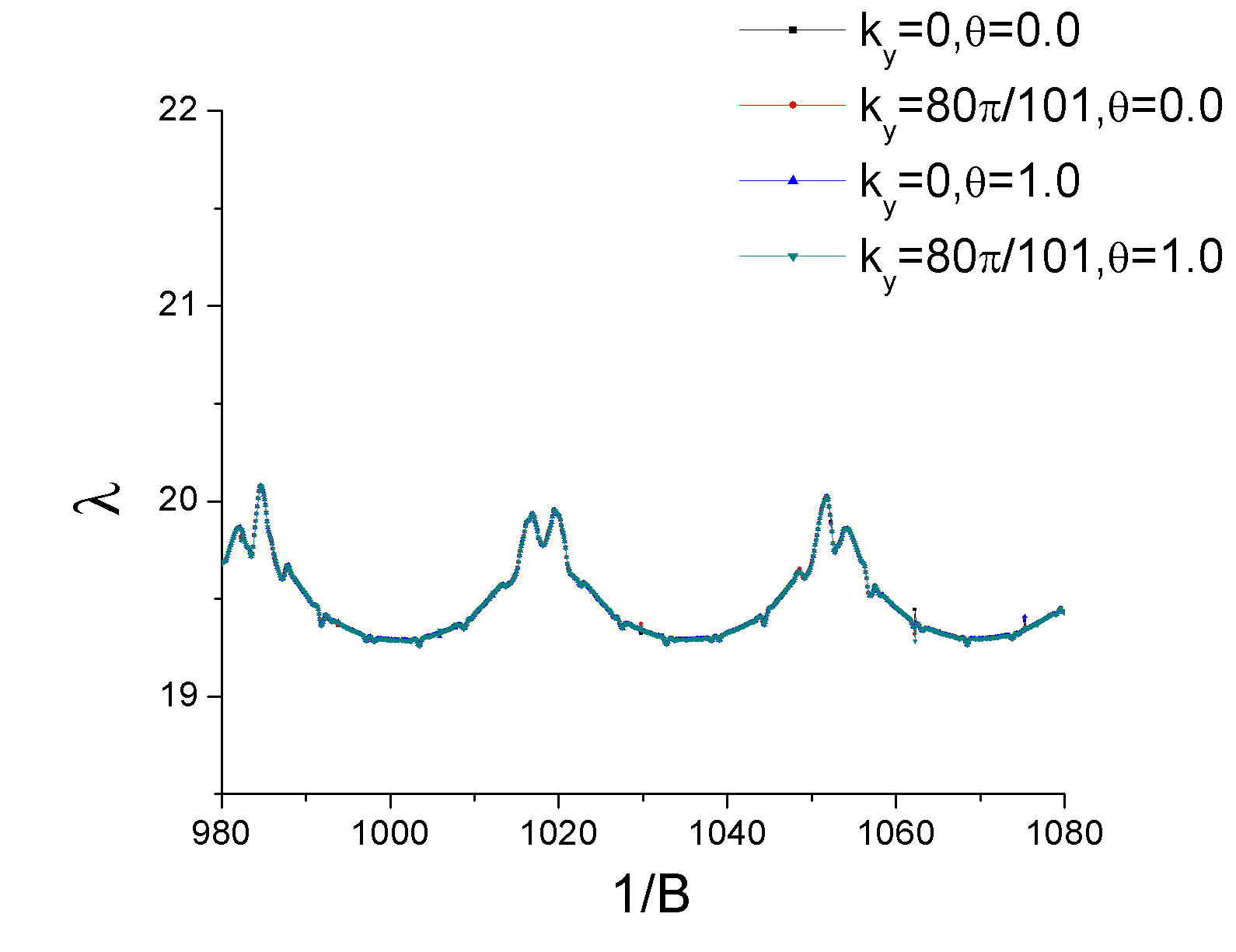}
\includegraphics[scale=0.155]{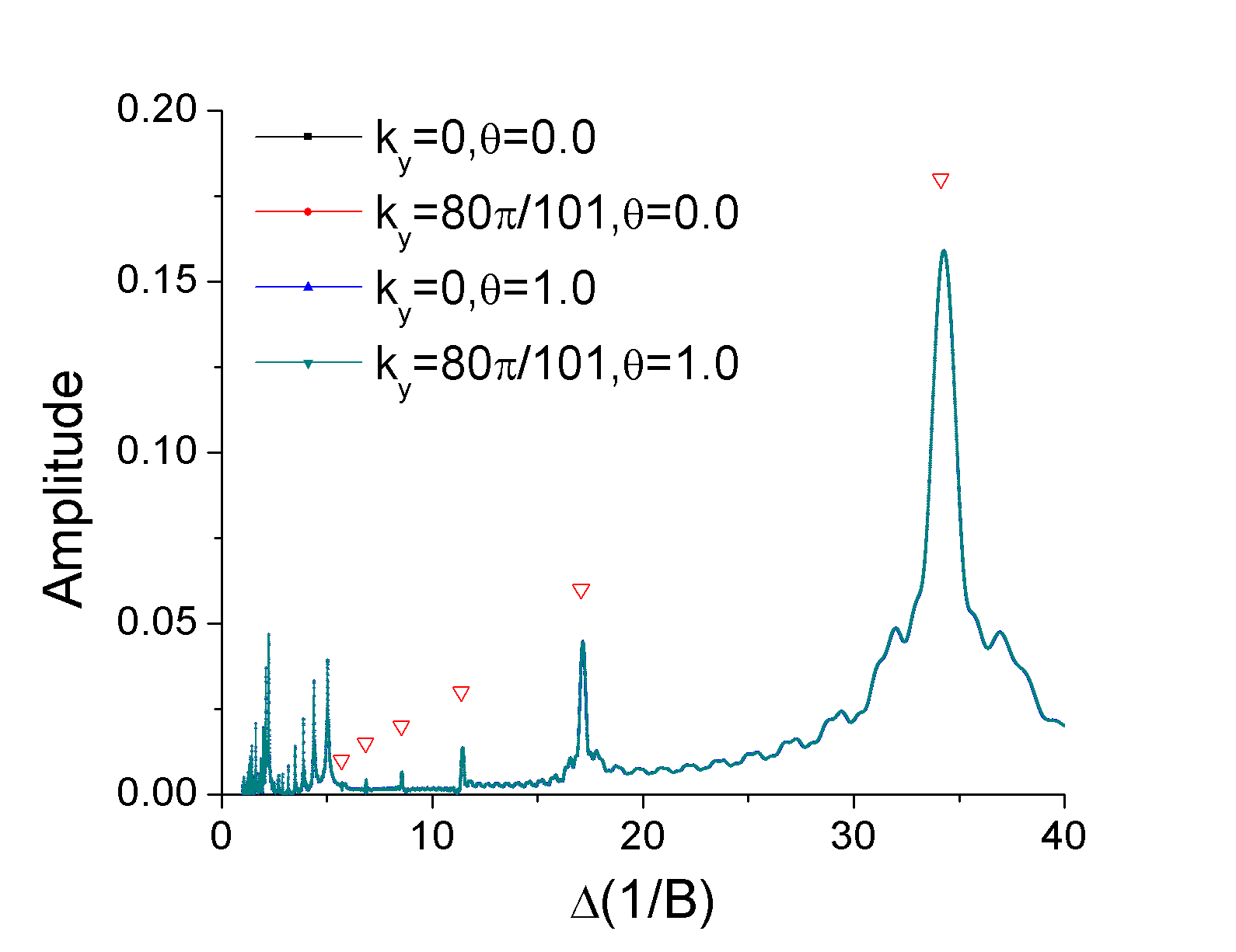}
\protect\caption{The localization length $\lambda$ versus the
inverse magnetic field $1/B$ for a system of length
$L=7.2\times10^{6}$, $\mu=-0.2$ and an ICDW with $Q=2.0$ and
$V=0.16$ over various ranges of the magnetic field. The last figure
is the Fourier transform of the QOs for the range
$1/B\in[280,1400]$. The red triangles mark the anticipated period
and corresponding higher harmonics from lowest order perturbation
theory. Different choices of $k_y$ and $\theta$ are shown in
different colors and their results clearly collapse onto the same
curve.} \label{fig5}
\end{centering}
\end{figure}

\begin{figure}
\begin{centering}
\includegraphics[scale=0.155]{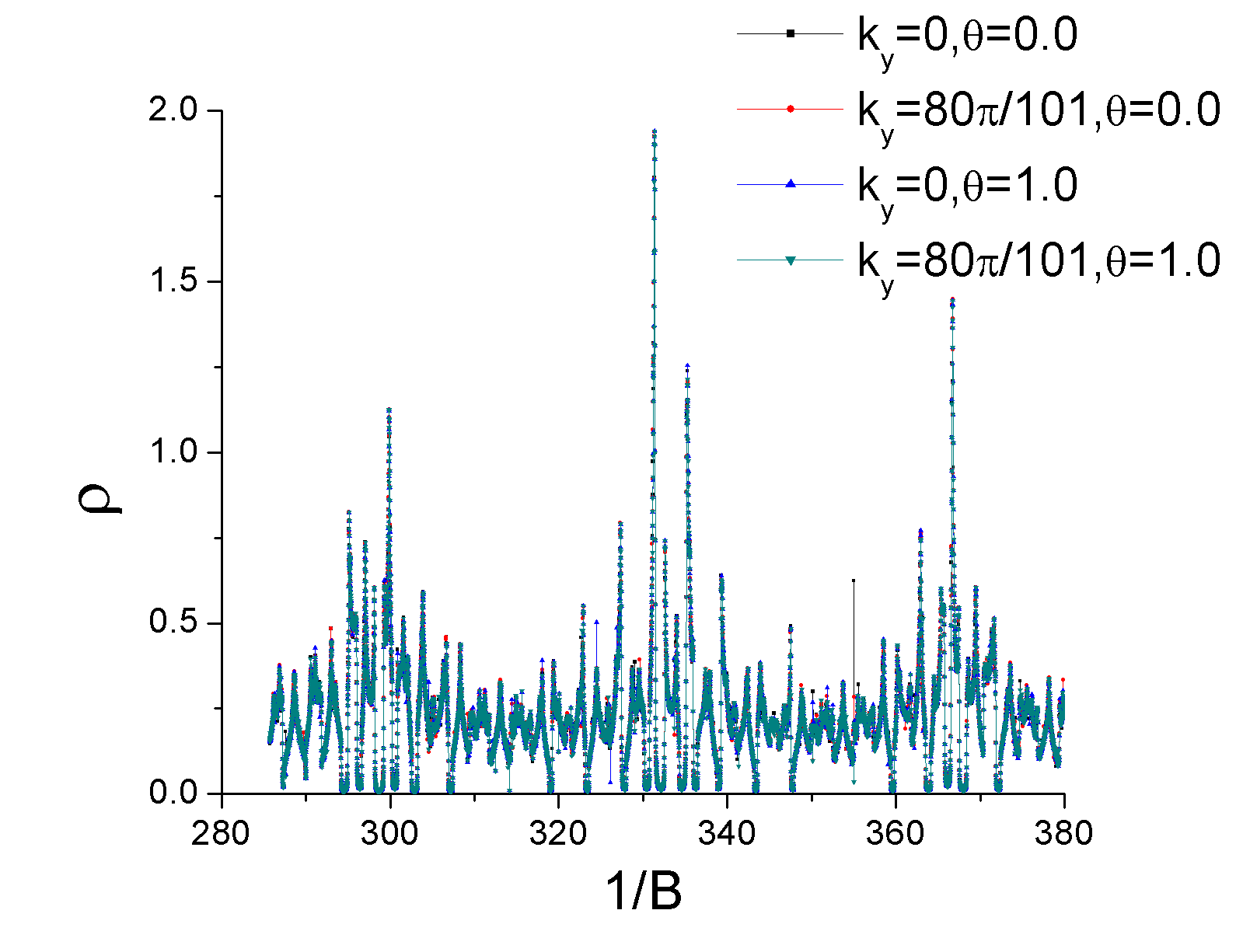}
\includegraphics[scale=0.155]{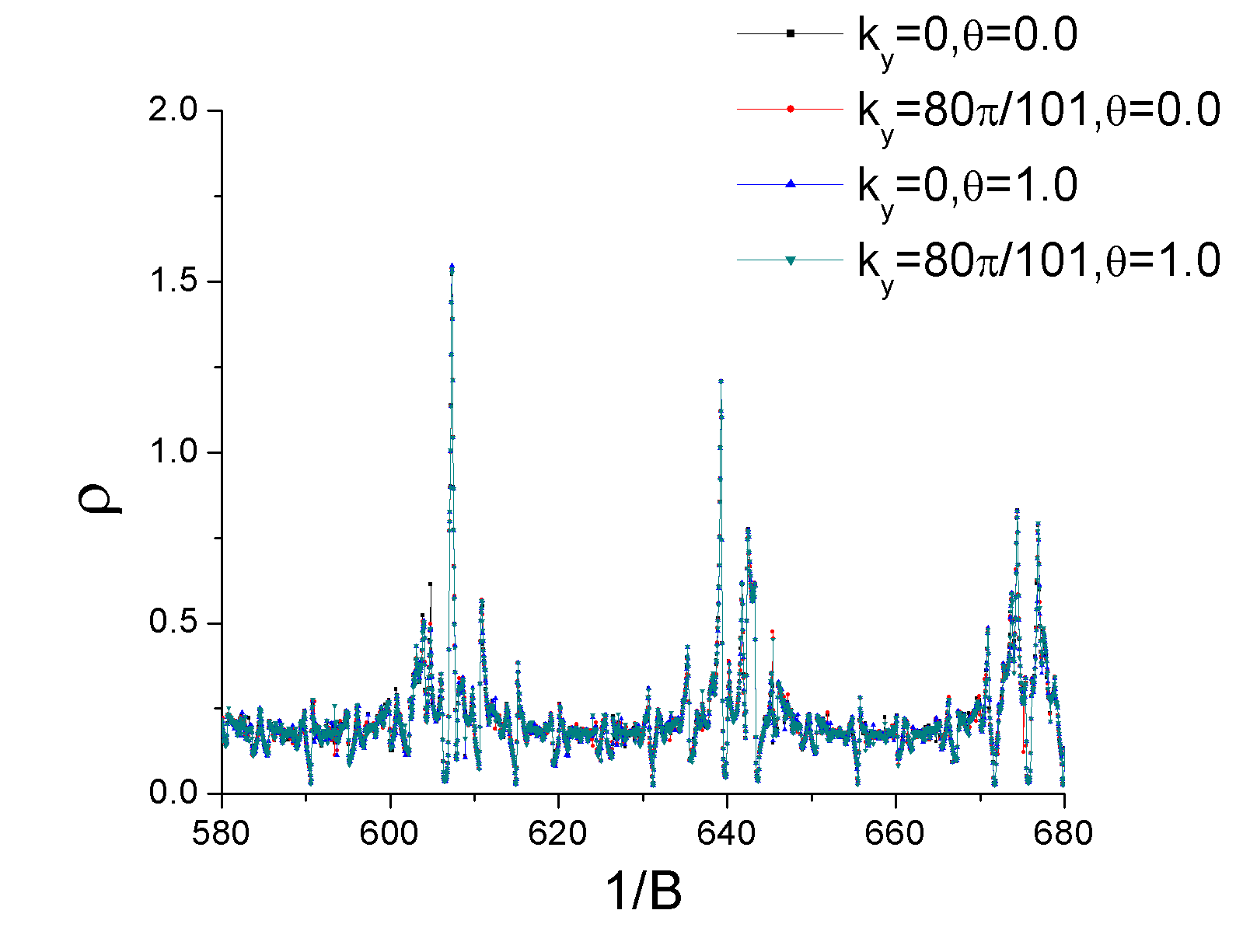}
\includegraphics[scale=0.155]{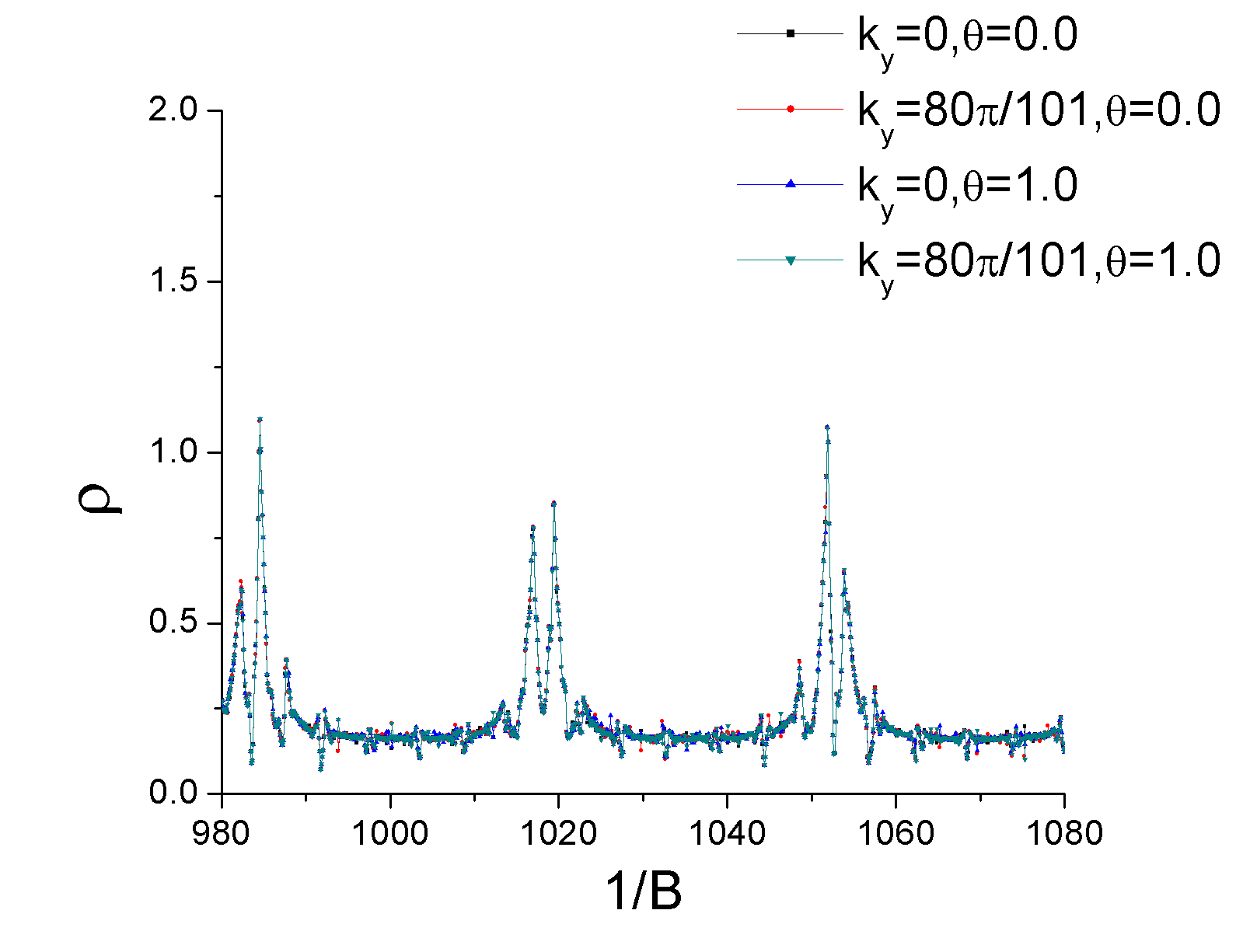}
\includegraphics[scale=0.155]{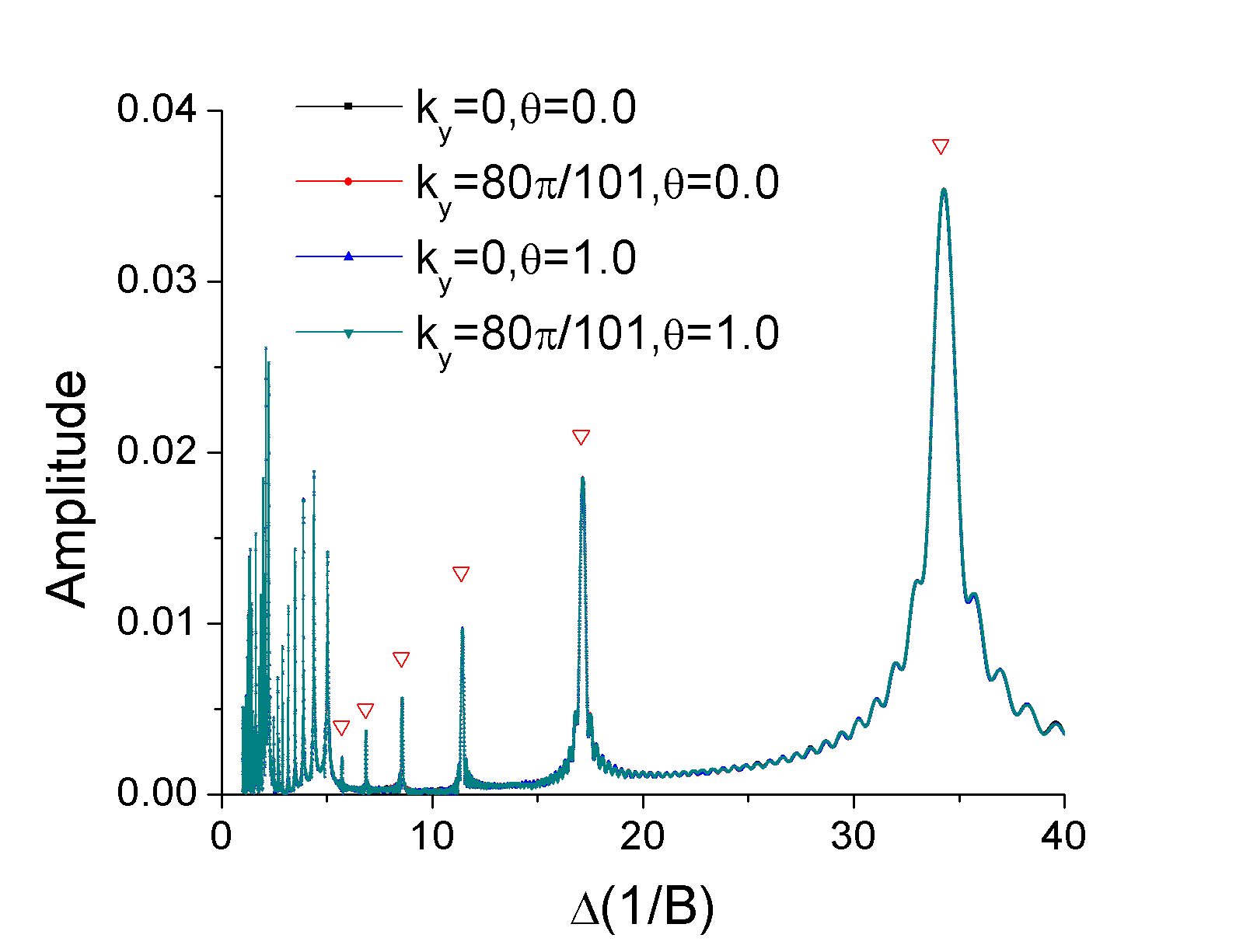}
\protect\caption{The DOS $\rho$ versus the inverse magnetic field
$1/B$ for a system of length $L=4\times10^{6}$, $\mu=-0.2$ and an
ICDW with $Q=2.0$ and $V=0.16$ over various ranges of the magnetic
field. The last figure is the Fourier transform of the QOs for the
range $1/B\in[280,1400]$. The red triangles mark the anticipated
period and corresponding higher harmonics from lowest order
perturbation theory. Different choices of $k_y$ and $\theta$ are
shown in different colors; the results clearly collapse onto the
same curve.} \label{fig6}
\end{centering}
\end{figure}

To have a secure basis for our discussion, we begin by applying the
same numerical technique described above to the incommensurate case.
As far as we are aware of, this is the first numerical solution of
the fully incommensurate problem. To begin with, we consider a CDW
with  wave vector $Q=2=2\pi(1/\pi)$ and potential $V=0.16$ with
$\mu=-0.2$; the results are shown in Fig. \ref{fig5} (localization
length $\lambda$) and Fig. \ref{fig6} (DOS $\rho$). For better
convergence and accuracy we work with systems of size
$L=7.2\times10^{6}$ for the localization length and
$L=4\times10^{6}$ for the DOS calculations. We find QOs with period
corresponding to the original (unreconstructed) Fermi surface
$\Delta\left(1/B\right)=2.245$ which are dominant when the field is
not too small, while another with $\Delta\left(1/B\right)=34.271$
becomes increasingly prominent at lower fields; each of these
fundamentals is accompanied by a complicated spectrum of higher
harmonics. In lowest order perturbation theory with a momentum
transfer of $\pm Q$, the area enclosed by the reconstructed Fermi
pocket (illustrated schematically as the light yellow patch in Fig.
\ref{fig2}) is $S_k=2.931\% S_{BZ}$ of the original Brillouin zone,
which corresponds to a period of $\Delta\left(1/B\right)=34.12$ as
we have marked in Fig. \ref{fig5} and \ref{fig6} together with
corresponding higher harmonics.

As expected, the above results are independent of the specific
choices of $k_y$ and $\theta$.

\begin{figure}
\begin{centering}
\includegraphics[scale=0.3]{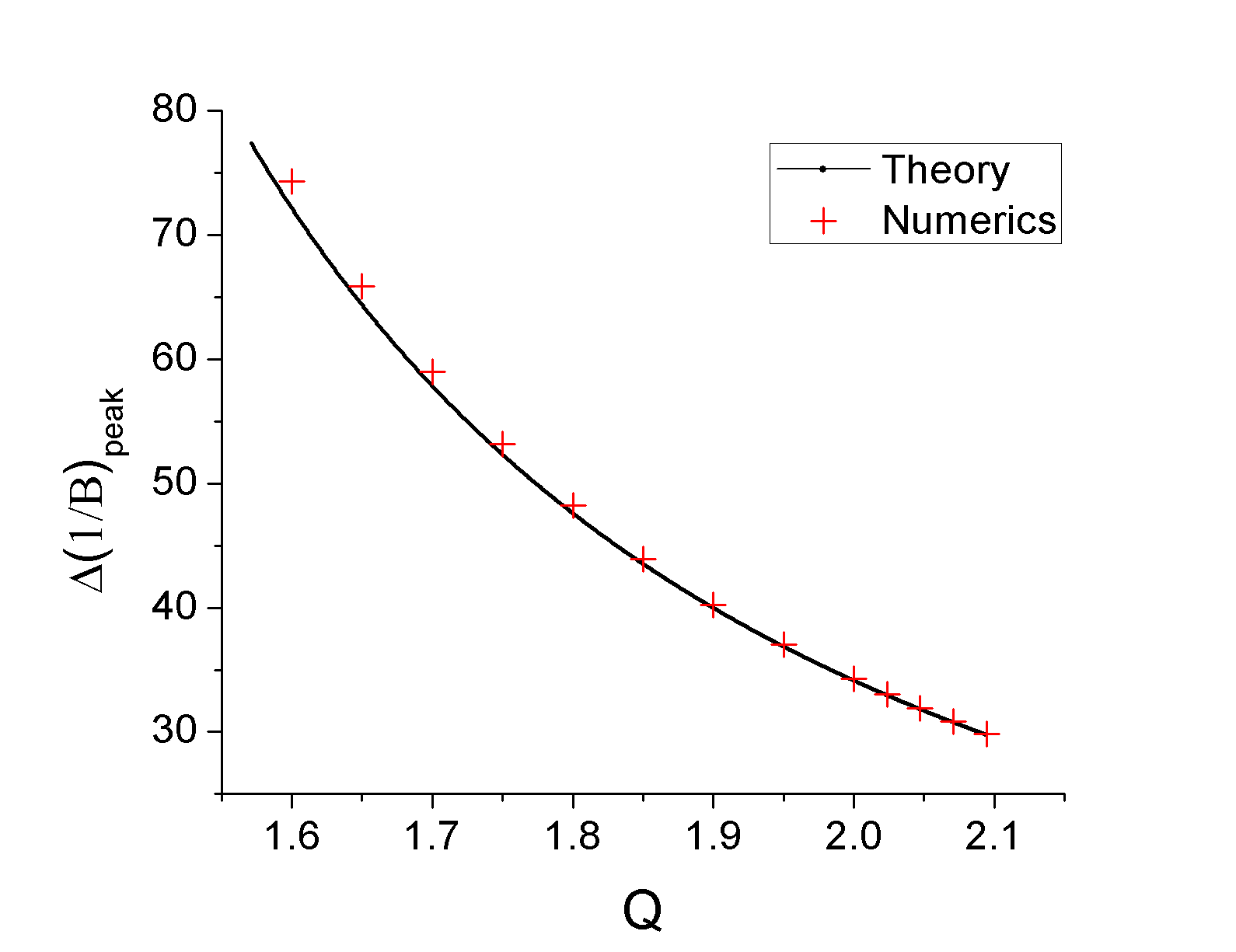}
\protect\caption{Red crosses: The calculated QOs characteristic
periods versus the ICDW wave vector $Q$ with $V=0.16$, $\mu=-0.2$
and $L=2\times10^5$. The black solid curve is the theoretical values
derived from the area of the first-order reconstructed Fermi
surfaces.} \label{fig7}
\end{centering}
\end{figure}

To further test the perturbative approach, we repeated the above
calculations with a series of ICDWs with wave vectors in the range
$Q\in(2\pi/4,2\pi/3)$ and with fixed values of $V$ and $\mu$. The
main oscillation periods obtained from the peak locations in the
Fourier transformed spectra are indicated by the crosses in Fig.
\ref{fig7}. For comparison, we also include the theoretical QO
period corresponding to the pocket sizes predicted from the lowest
order perturbation. The consistency is remarkable.

\subsubsection{Breakdown of perturbation theory as $B\to 0$}

\begin{figure}
\begin{centering}
\includegraphics[scale=0.3]{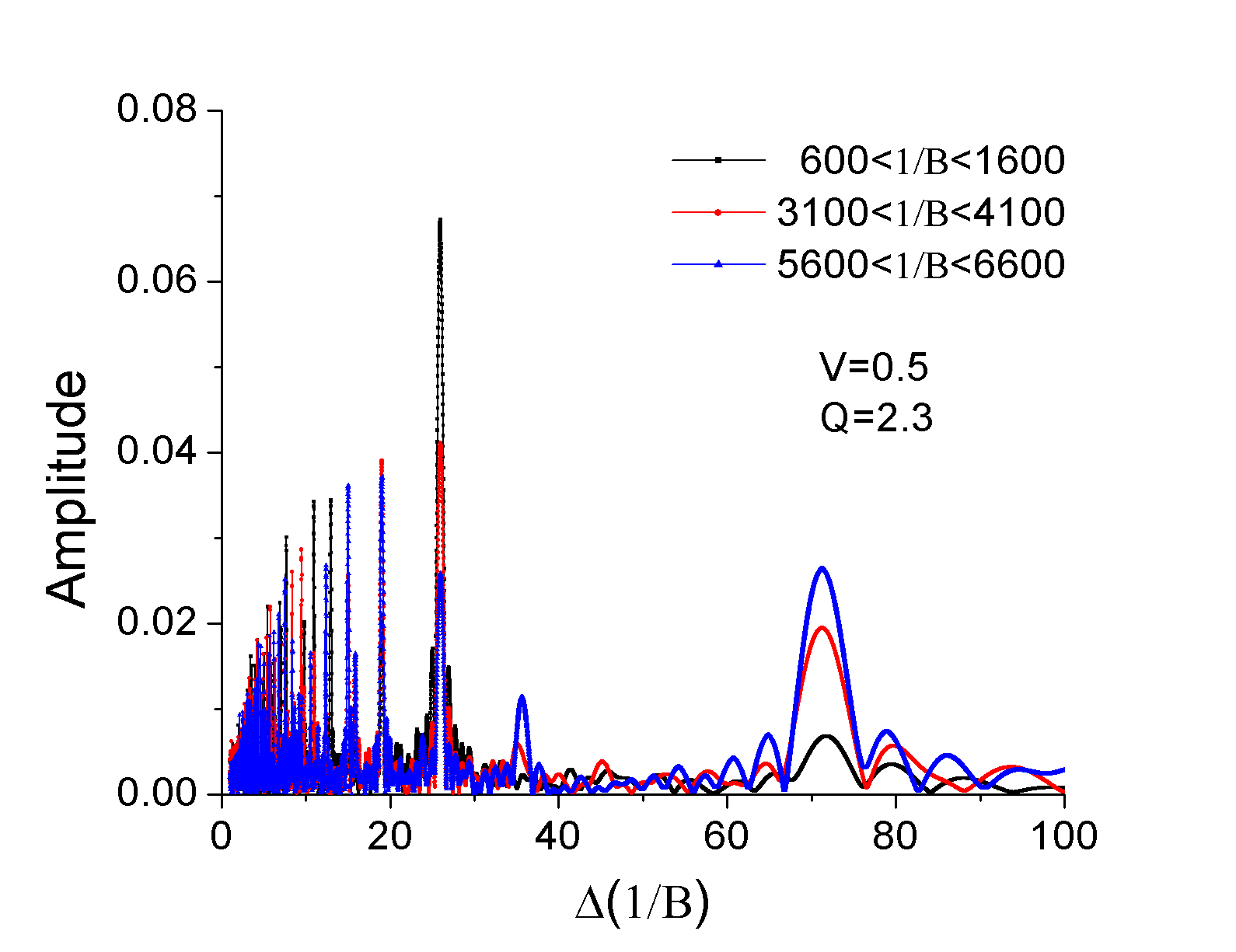}
\includegraphics[scale=0.3]{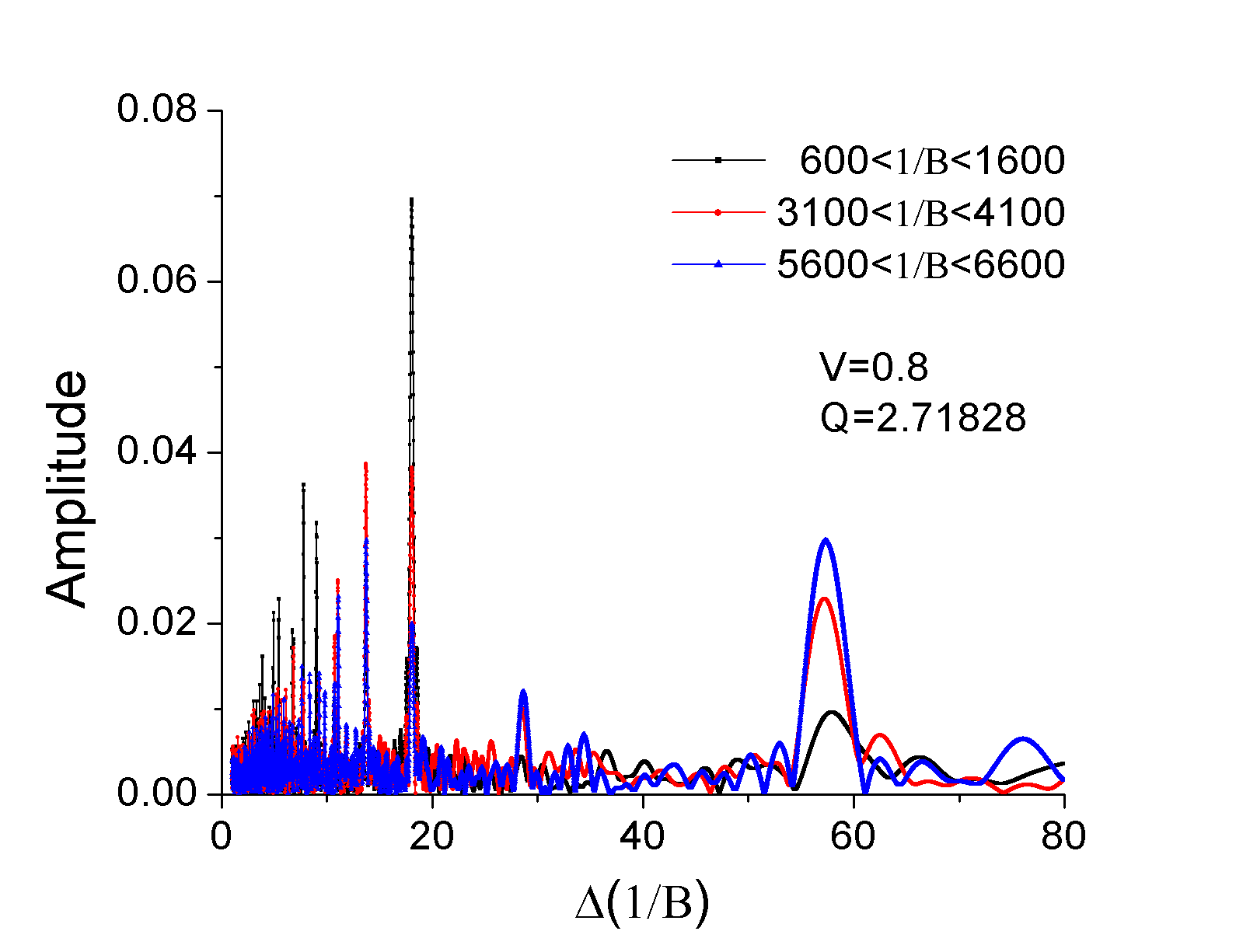}
\protect\caption{The Fourier transformations of the QOs of the DOS
$\rho$ with parameters $V=0.5$, $Q=2.3$, $\mu=-0.2$ (upper panel)
and $V=0.8$, $Q=2.71828$, $\mu=-0.2$ (lower panel) over various
ranges of the magnetic field. The system size is $L=2\times10^5$.}
\label{fig19}
\end{centering}
\end{figure}

We have anticipated that the finite-order perturbative results
inevitably break down for small enough magnetic fields. Since any
gaps generated in $n^{th}$ order perturbation theory are of order
$W(V/W)^n$, the characteristic magnetic field scale above which
these gaps are eliminated by magnetic breakdown is expected to scale
with $V$ according to $B_n\sim V^{2n}$.  This leads to the cascade
of field ranges that has already been exhibited in the commensurate
case above. The difference is that, while in the case of
commensurability $q$ no new gaps are generated beyond $q^{th}$ order
perturbation theory, in the incommensurate case no termination of
the cascade is expected. For numerical experiments, it is simpler to
study this  working in a fixed range of magnetic fields and
increasing the magnitude of the CDW potential $V$, rather than by
going to still lower fields.

We thus calculate the QOs of the DOS for somewhat larger magnitude
of the CDW potential. The results for two illustrative examples:
$Q=2.3$, $V=0.5$ and $Q=2.71828$, $V=0.8$ are shown in Fig.
\ref{fig19}. We have kept the chemical potential $\mu=-0.2$ and
taken $L=2\times 10^5$ in these calculations. Different ranges of
the magnetic field are shown in different colors to emphasize the
cross-over between different regimes.

For $Q=2.3$, distinct points on the Fermi surface are connected by
$Q$ and $2Q$, so we expect the QOs to be determined principally by
the area of the unperturbed Fermi surface for high fields, $B >
B_1\sim (V/W)^2$, the first order perturbatively reconstructed Fermi
surface for $B_1 > B > B_2\sim (V/W)^4$, and the second order
reconstructed Fermi surface for $B_2 > B
> B_3\sim (V/W)^6$.  We do not report the data for the highest
fields, $B> B_1$ but in the upper panel of Fig. \ref{fig19} one can
observe a crossover from the dominant oscillations having period of
$\Delta(1/B)=25.97$ in the higher field range to $\Delta(1/B)=71.24$
in the lower; these are consistent with the expected frequencies
corresponding to a Fermi surface reconstructed to first and second
order in $V$, respectively. On the other hand, for $Q=2.71828$, the
scattering vector $2Q$ does not span the Fermi surface, so no new
gaps on the Fermi surface are opened in second order in perturbation
theory. However, scattering with momentum $Q$ and $3Q$ (modulo
$2\pi$) opens distinct gaps on the Fermi surface, and indeed the
expected crossover from a dominant oscillation period of
$\Delta(1/B)=18.07$ at relatively high fields to $\Delta(1/B)=57.32$
at somewhat lower fields is observed, in good agreement with the
expected periods from the perturbative construction.

We conclude from the numerical results for $V$ smaller than $W$ and
not too small magnetic fields that perturbation theory provides a
grossly satisfactory description of the structure of the QOs, and
that it also provides a vivid account of the manner in which the
perturbative results to any given order break down for low enough
magnetic fields.

\subsubsection{Absence of all QOs for large $V$}

\begin{figure}
\begin{centering}
\includegraphics[scale=0.3]{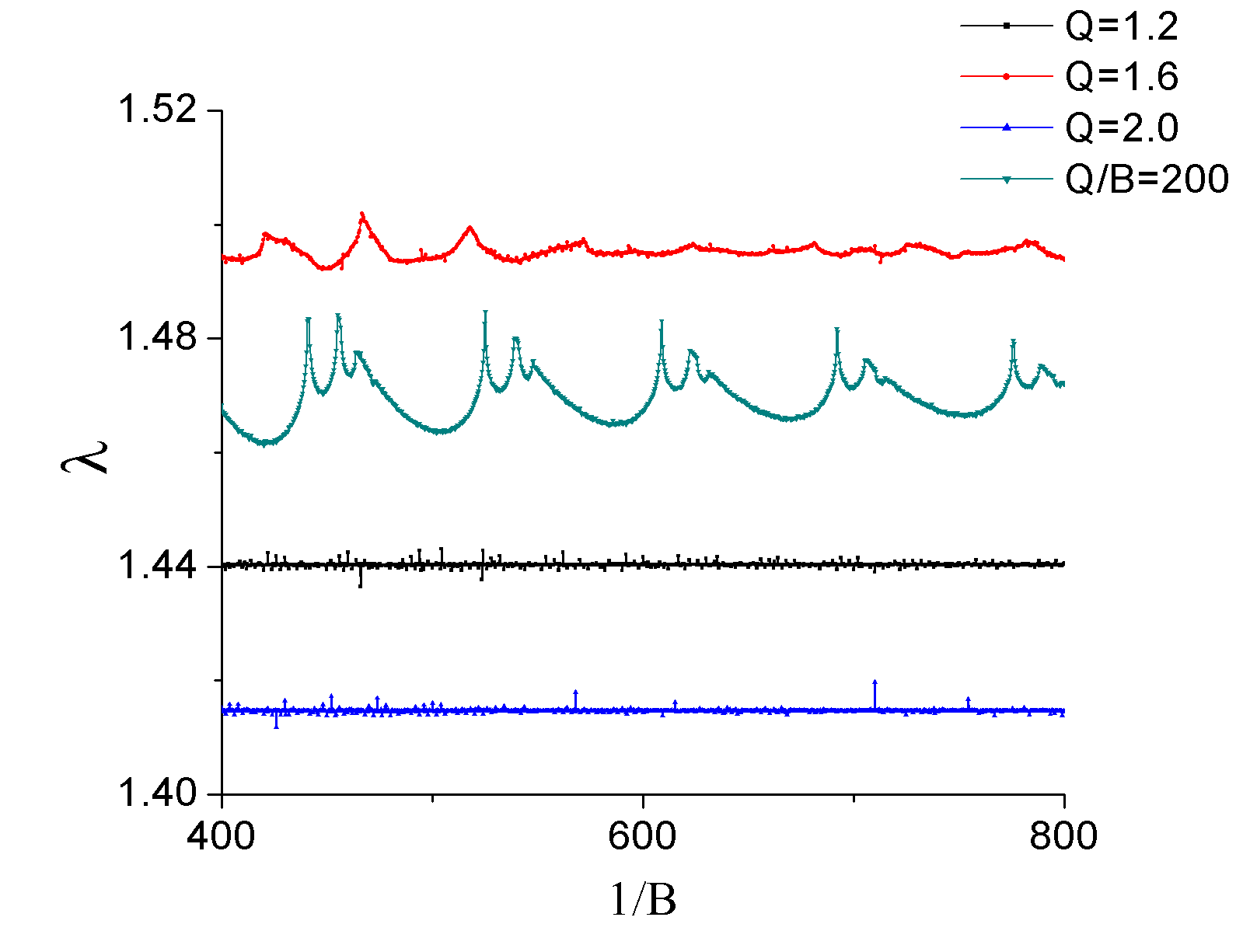}
\protect\caption{The localization length $\lambda$ in the presence
of a strong ICDW $V=1.6$, $\mu=-2.5$ and $Q=1.2$ (black), $Q=1.6$
(red) and $Q=2.0$ (blue). For comparison, we include the QOs of
localization length $\lambda$ with $V=1.6$, $\mu=-2.5$ and $Q/B=200$
for the same range of magnetic field $B$ (green) with
$L=7.2\times10^5$.} \label{fig14}
\end{centering}
\end{figure}

For large $V>W$, clearly perturbation theory is expected to have no
regime of validity. On the other hand, the behavior can be well
understood in a dual picture using the mapping of the problem onto
the properties of a three-dimensional crystal with no CDW but in the
presence of a tilted effective magnetic field, $\vec{B}^{\hskip2pt
\rm eff}=\left[ B\hat{z}+Q\hat{y}/2\pi\right]$, as we will discuss
in Sec. \ref{dual}. In particular, we will show that for $V$ larger
than a characteristic magnitude $V_c$ (defined as the point at which
a Lifshitz transition occurs in the band-structure of the dual
three-dimensional crystal), there are no well defined QOs at all!

To illustrate the basic phenomenology, we show in Fig. \ref{fig14}
the localization length as a function of $B$ in the presence of a
reasonably strong CDW potential with $V=1.6$, $\mu=-2.5$ and
$Q=1.2$, $1.6$, and $2.0$. Also shown in the figure is the
corresponding curve in the unphysical case in which the CDW ordering
vector varies in proportion to the magnetic field as $Q=\alpha B$,
with the constant $\alpha=200$ chosen so that $Q$ is of order $1$ in
the range of fields represented in the figure. Manifestly, the three
curves with fixed $Q$ show no periodic oscillations at all, and
indeed relatively weak magnetic field dependence of any sort.
However, when we vary $Q$ in proportion to $B$ we do see perfectly
periodic QOs, albeit with a rather unusual harmonic content. These
features are easily understood in the dual three-dimensional
picture, which we now discuss.

\section{The dual three-dimensional view of quantum oscillations}
\label{dual}

\begin{figure}
\begin{centering}
\includegraphics[scale=0.35]{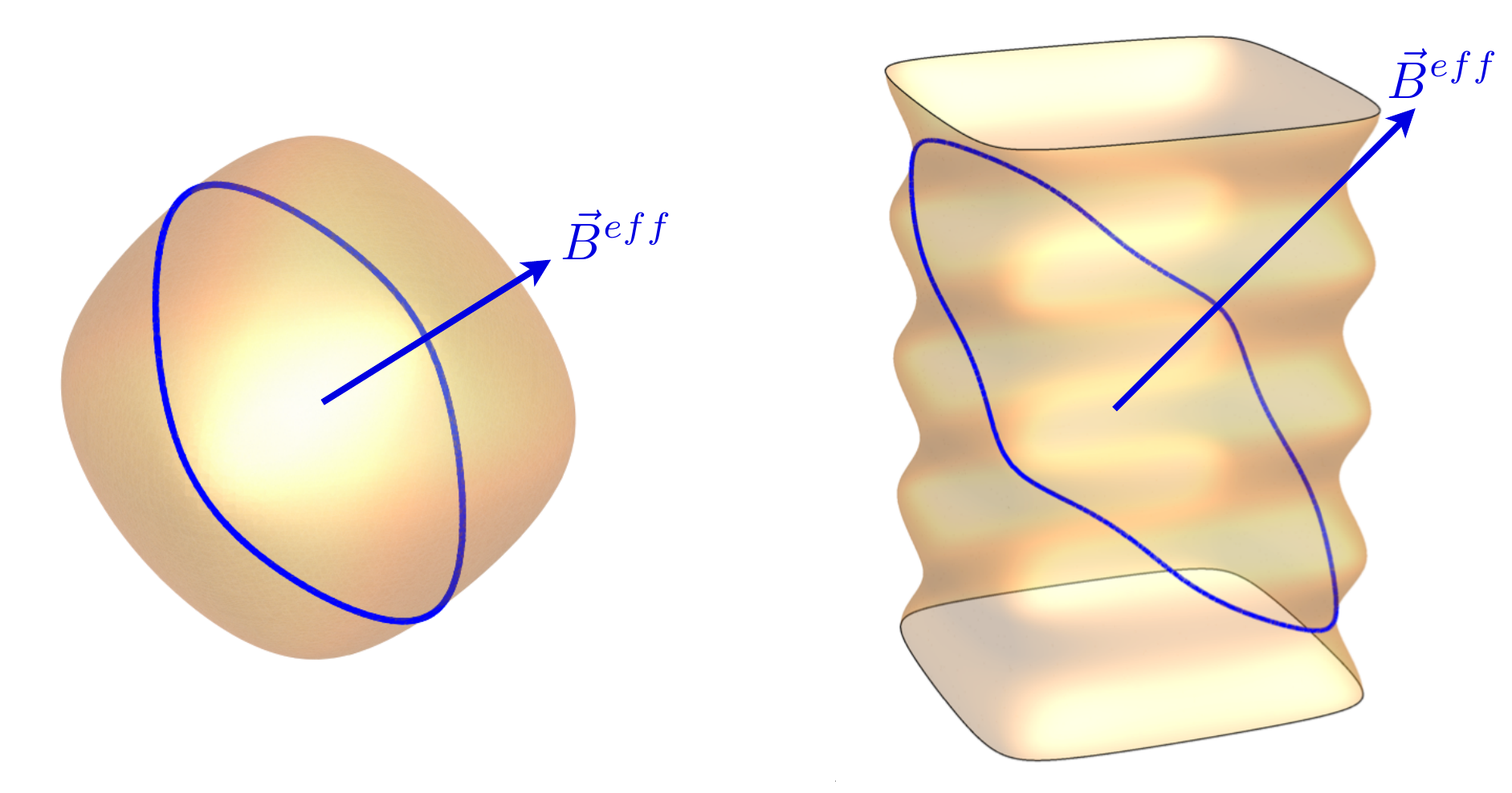}
\protect\caption{Zero-field Fermi surfaces of the three-dimensional
effective theory for (left) a strong ICDW with $V=1.6$, $\mu=-2.5$
and (right) a weak ICDW with $V=0.16$, $\mu=-0.2$. The cross
sections perpendicular to $\vec{B}^{\hskip2pt \rm eff}$ are shown as
the blue lines.}\label{fig9}
\end{centering}
\end{figure}

The advantage of the dual three-dimensional representation of the
problem is that it allows us to treat the CDW potential $V$
non-perturbatively, by incorporating it into the band-structure.
Then provided the magnitude of $\vec B^{\hskip2pt \rm eff}
=\left(B\hat{z}+Q\hat{y}/2\pi\right)$ is sufficiently small (in a
sense we will define below) we can again treat the electron dynamics
semiclassically, although now on a three-dimensional Fermi surface.

\subsection{Numerical tests of duality}

Although the duality is exact in the thermodynamic limit, we begin
by performing numerical tests to establish the accuracy of the
approach for the system sizes used in this study. For this purpose,
we use the same numerical approach to solve the one-dimensional
problem with a doubly incommensurate potential, and compare the
results to expectations based on the three-dimensional effective
theory. This is made simpler if we consider the unphysical situation
in which both $B$ and $Q$ are varied simultaneously, while keeping
the ratio $B/Q$ fixed, so that the direction of the effective
magnetic field $\vec{B}^{\hskip2pt \rm
eff}=\left(B\hat{z}+Q\hat{y}/2\pi\right)$ remains fixed.
Semiclassically, QO frequencies are simply given by the maximum and
minimum cross-section areas of the three-dimensional Fermi surface
perpendicular to $\vec{B}^{\hskip2pt \rm eff}$, and the numerical
results in Fig. \ref{fig14} confirm QOs are present, even for large
$V$.

\begin{figure}
\begin{centering}
\includegraphics[scale=0.3]{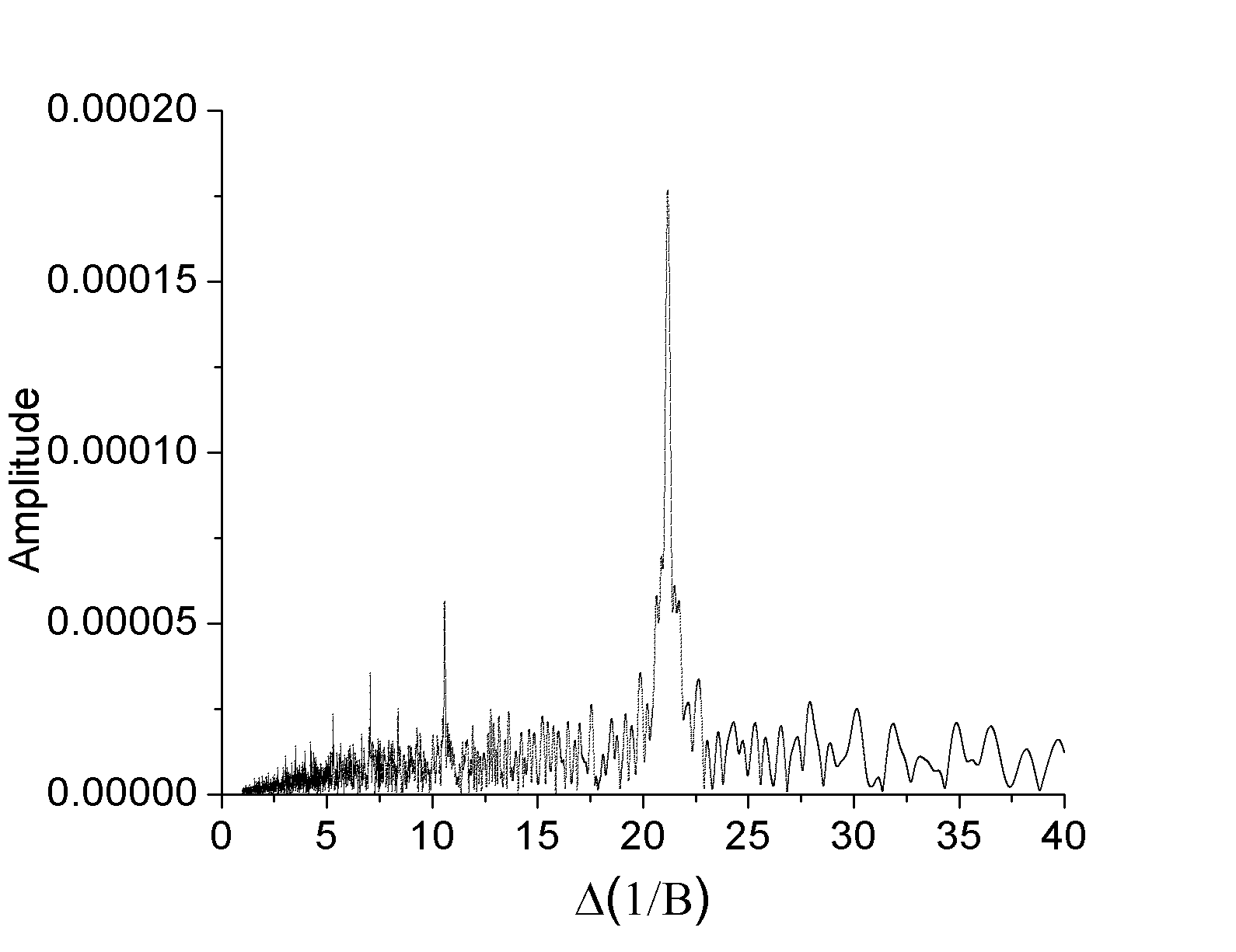}
\protect\caption{Fourier transformations of the QOs of the DOS
$\rho$ with a strong ICDW $V=1.6$ and $\mu=-2.5$ and $Q/B=50$ over a
range of $1/B\in[500,2500]$. A clear peak is observed at
$\Delta(1/B)=21.175$ as well as its higher harmonics. The system
size is $L=7.2\times10^5$.} \label{fig10}
\end{centering}
\end{figure}

\emph{Strong CDW} ($V=1.6$, $\mu=-2.5$): In this case, the CDW
potential is sufficiently large that the Fermi surface in the
three-dimensional effective theory is closed. Fig. \ref{fig9} (left
panel) provides an illustration of the three-dimensional Fermi
surface structure, which is topologically equivalent to a sphere and
closed in the $\hat{z}$ direction. We numerically obtain QOs of the
DOS for $Q/B=50$ over a range of $1/B\in[500,2500]$ on system size
of length $L=7.2\times10^{5}$, and the resulting Fourier transforms
is shown in Fig. \ref{fig10}. A clear single peak is observed at the
expected position as predicted from the Fermi surface maximum
cross-section area $S_{BZ}/S_{k}=21.174$ perpendicular to
$\vec{B}^{\hskip2pt \rm eff}$.

\begin{figure}
\begin{centering}
\includegraphics[scale=0.3]{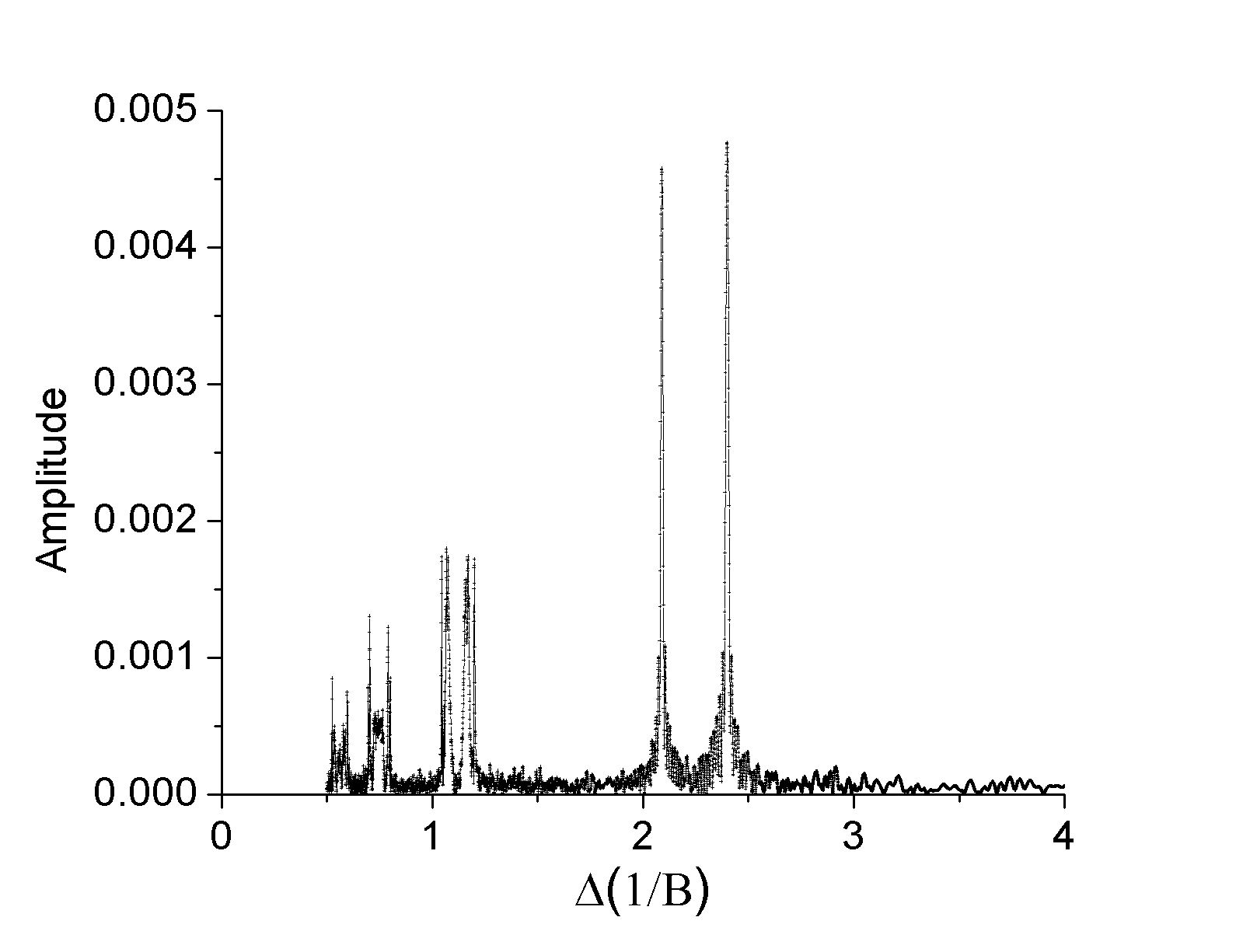}
\includegraphics[scale=0.15]{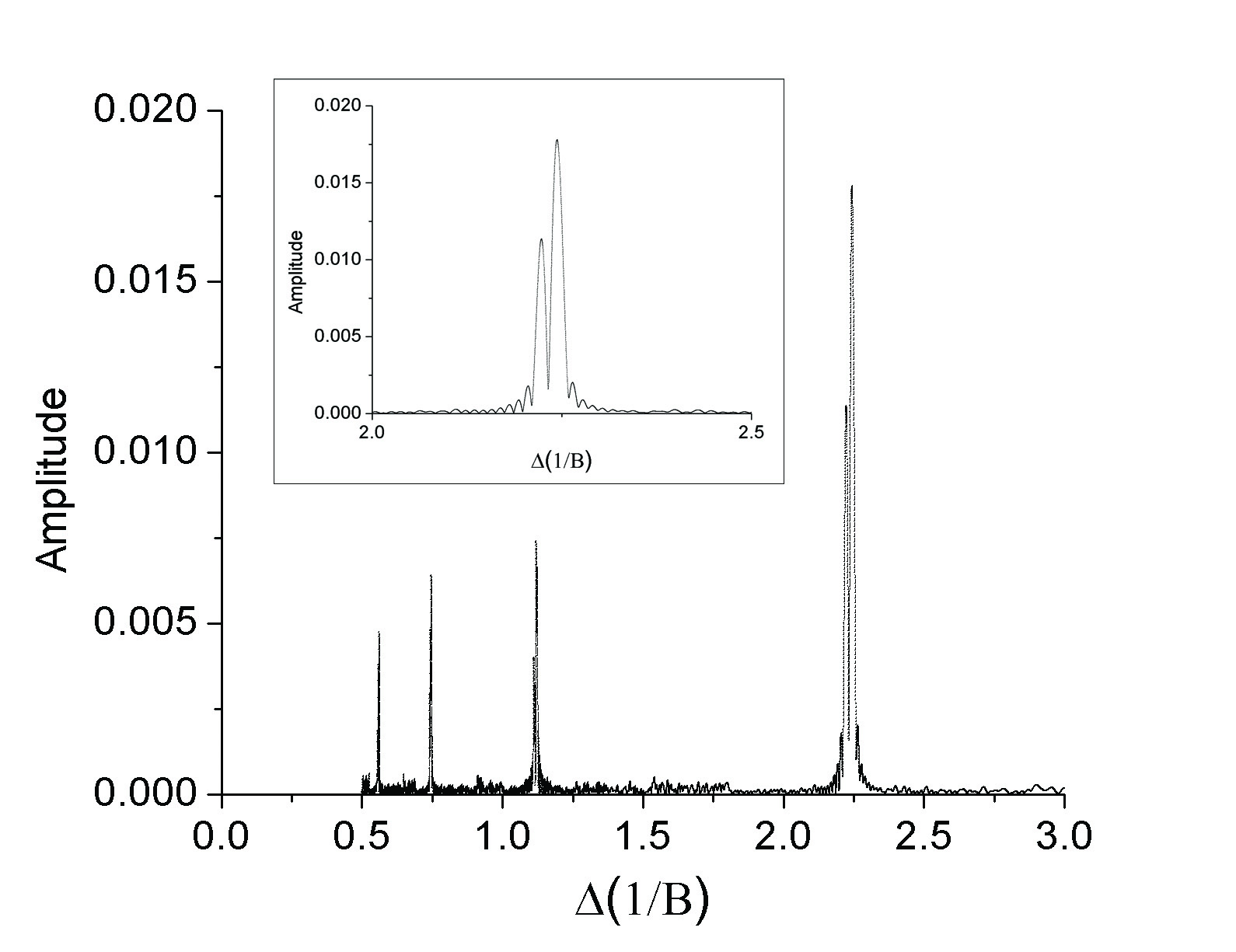}
\protect\caption{Fourier transformations of the QOs of the DOS
$\rho$ 
for a weak CDW $V=0.16$ and $\mu=-0.2$ over a range of
$1/B\in[1000,1400]$. A clear double-peak structure is observed.
Upper panel: $Q/B=2$, the peaks are at $2.089$ and $2.399$ as well
as their higher harmonics; Lower panel: $Q/B=20$, the peaks are at
$2.2233$ and $2.239$ as well as their higher harmonics. The inset is
an enlargement over the range of the double peaks. The system size
is $L=7.2\times10^5$.} \label{fig11}
\end{centering}
\end{figure}

\emph{Weak CDW} ($V=0.16$, $\mu=-0.2$): In this case, the Fermi
surface is just slightly modified from the stacking of the model's
two-dimensional Fermi surfaces and nearly cylindrically shaped, see
right panel of Fig. \ref{fig9}. The periods of QOs are now dominated
by both the maximum and the minimum Fermi surface cross-section
areas perpendicular to $\vec{B}^{\hskip2pt \rm eff}$. In particular,
we calculate the QOs of the DOS over a range of $1/B\in[1000,1400]$
on system size of length $L=7.2\times 10^{5}$, and the resulting
Fourier transforms for the cases of $Q/B=2$ and $Q/B=20$ are shown
in Fig. \ref{fig11}. A clear double peak structure is observed in
both calculations. Again, the peak positions are consistent with the
theory deduction from maximal and minimal Fermi surface cross
section areas $S_{BZ}/S_{k}^{MAX}=2.399$ and
$S_{BZ}/S_{k}^{MIN}=2.089$ for the $Q/B=2$ case and
$S_{BZ}/S_{k}^{MAX}=2.2468$ and $S_{BZ}/S_{k}^{MIN}=2.2226$ for the
$Q/B=20$ case, respectively.

\subsection{Results}

In the physical case, $Q$ is generally expected to be approximately
independent of magnetic field. (Although note that there are cases,
such as the famous field-induced SDW states in (TMTSF)$_2$ClO$_4$,
in which $Q$ is strongly $B$ dependent.\cite{chaikin}) In this case,
both the magnitude and direction of $\vec B^{\hskip2pt \rm eff}$
varies as a function of $B$. In terms of the three-dimensional
semiclassical trajectories, this means that the orientation of the
orbits, and hence extremal cross-section areas $S^{\perp}_{k}(B)$,
vary as a function of the applied magnetic field $B$, in addition to
its usual dependence on the band structure parameters $t$, $V$, and
$\mu$. Since we are typically most interested in the small $B$
limit, the most important cases are those in which $B^{\hskip2pt \rm
eff}$ is only tilted slightly away from the $\hat y$ directions, and
hence the semiclassical orbits lie close to the $x-z$ plane. Maxima
in the semiclassical DOS are therefore found when
\begin{equation}
\frac{1}{\sqrt{B^2 + (Q/2\pi)^2}} = \left(n + \frac{1}{2}\right)
\frac{2\pi}{S^{\perp}_{k}(B)}\label{eq:3dsemiclassic}
\end{equation}

From this equation it is clear that QOs which are perfectly periodic
in $1/B$ are no longer possible for generic three-dimensional Fermi
surfaces (i.e. generic cross-section areas $S^{\perp}_k(B)$). There
is also a qualitative difference in the semiclassical trajectories
for $V < V_c$, for which the Fermi surface forms a warped cylinder
which is open in the $\hat z$ direction as shown in the right panel
of Fig. \ref{fig9}, vs. $V>V_c$, in which case the Fermi surface is
closed in the $\hat z$ direction, as shown in the left panel of the
figure. Here $V_c(\mu)$ is the critical value of $V$ at which a
Lifshitz transition occurs at which the three-dimensional Fermi
surface changes its topology.

\begin{figure}
\begin{centering}
\includegraphics[scale=0.3]{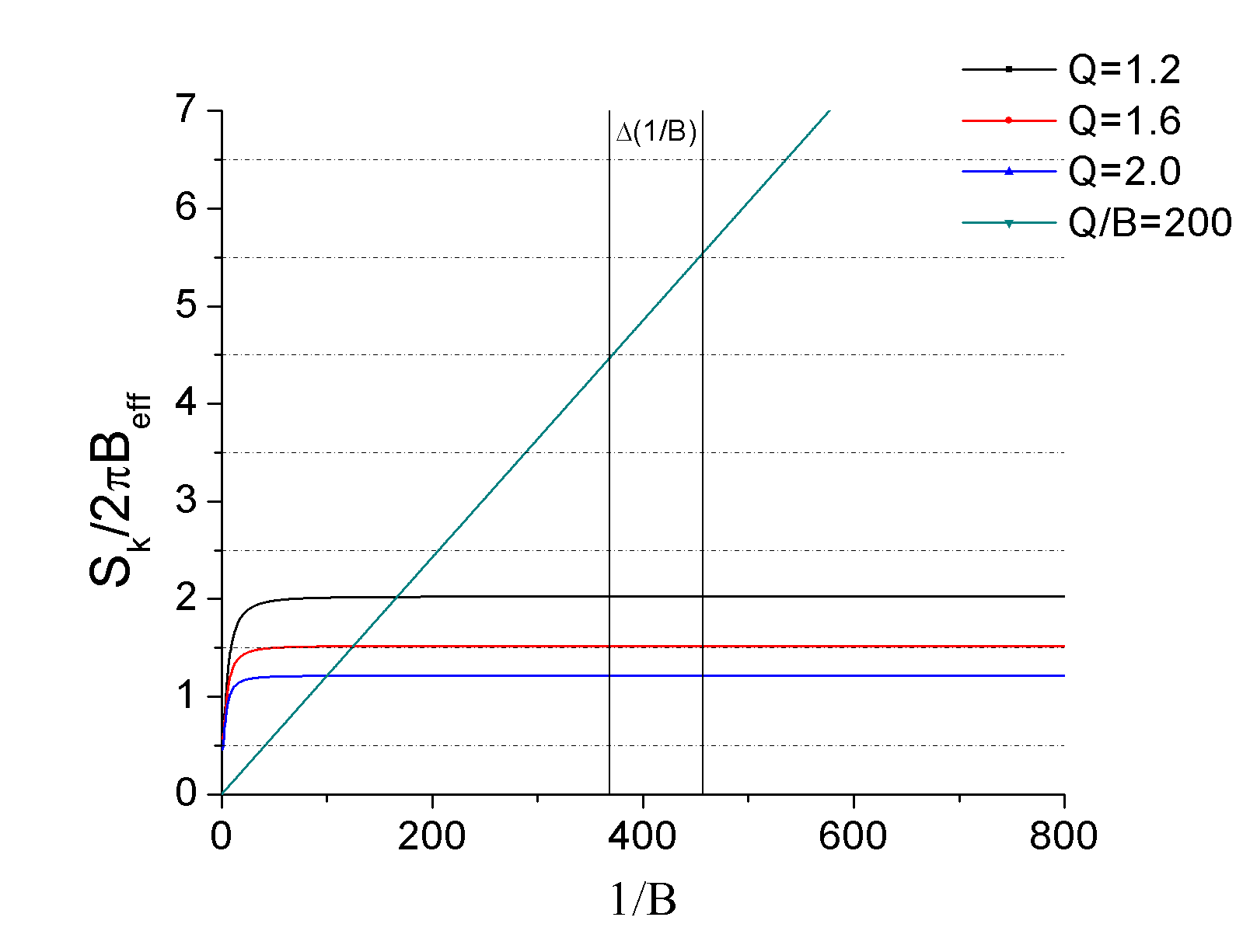}
\protect\caption{The value of the orbital phase factor $S_{k}/2\pi
B_{\hskip2pt \rm eff}$ over a range of magnetic field $B$:
$1/B\in[0,800]$ for a strong ICDWs with parameters $V=1.6$,
$\mu=-2.5$ and $Q=1.2$ (black), $Q=1.6$ (red), $Q=2.0$ (blue),
respectively. $S_k$ is the maximal area of cross-sections
perpendicular to $\vec B_{\hskip2pt \rm eff}$ for each given $B$.
Results for fixed $Q/B=200$ are also shown for comparison (cyan).
Peaks in the DOS occur whenever these curves cross half-integer
values, shown in the figures as the horizontal dashed lines.}
\label{fig13}
\end{centering}
\end{figure}

\subsubsection{Absence of QOs for $V > V_c$}
So long as $V>V_c$, there is a well defined maximal cross-sectional
area of Fermi surface contours perpendicular to $\hat y$. Moreover,
for small $B$, the maximal cross-sectional area perpendicular to
$\hat B^{\hskip2pt \rm eff}$, $S_{k}^{\perp}\lesssim
S_{k,MAX}^{\perp}$, is close to and bounded by the maximum area for
cross sections perpendicular to $\hat y$. Therefore, even if $B$ is
changed by an arbitrarily large factor, so long as $B \ll Q$,
neither the magnitude of $\vec B^{\hskip2pt \rm eff}$, nor the value
of $S_{k}^{\perp}$ changes substantially, and hence little variation
is expected of the electronic structure. For example, we show in
Fig. \ref{fig13} the value $S_{k}/2\pi B_{\hskip2pt \rm eff}$ over a
range of $1/B$ with parameters $V=1.6$, $\mu=-2.5$ and $Q=1.2$,
$1.6$, and $2.0$. These curves do not intersect the half integers
and are constrained to the quantum limit with only few Landau levels
occupied for reasonably small $B$, thus explaining the absence of
QOs reported in Fig. \ref{fig14}. (The ``ghost'' of QOs in the
$Q=1.6$ case can be attributed to the fact that $S_{k}/2\pi
B_{\hskip2pt \rm eff}$ passes close to a half-integer value, where
due to its vicinity to resonance the physical quantities are
sensitive to small variation of the orbital phase factor). This can
be contrasted with the case with a fixed $Q/B=200$, where $S_{k}$ is
invariant and $S_{k}/2\pi B_{\hskip2pt \rm eff}$ is linear in
$B_{\hskip2pt \rm eff}$, producing periodic QOs as a function of
$1/B$.

Interestingly, for the same $V=1.6$, $\mu=-2.5$ but a nearby
commensurate CDW with $Q=\pi/2$, the model can be considered in the
conventional framework of Sec. \ref{sec:breakdowns}, and is a metal
with one piece of closed Fermi surface. As expected, numerical
calculations show clear and perfectly periodic QOs. Yet, we have
shown that for a nearby incommensurate wave vector $Q=1.6$ in Fig.
\ref{fig14}, there are no well-defined QOs. We emphasize that
treating the ICDW as an approximately commensurate CDW can be
grossly misleading.

\subsubsection{Breakdown of semiclassical approximation for $V< V_c$}

For small values of $V/W$, the perturbative picture is useful.
Nevertheless several new effects become apparent, even in this
limit, when viewed from the dual perspective.

\begin{figure}
\begin{centering}
\includegraphics[scale=0.3]{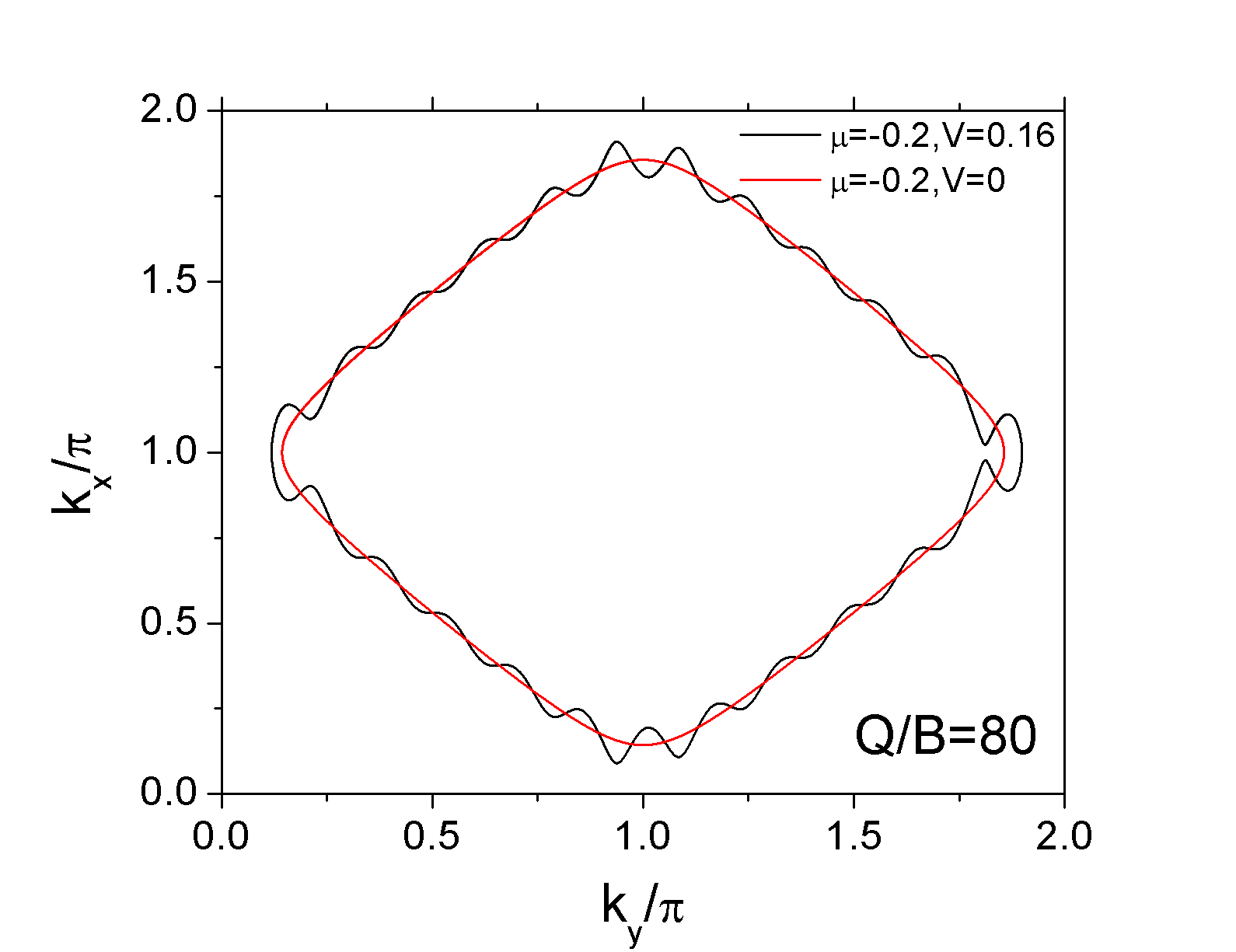}
\includegraphics[scale=0.15]{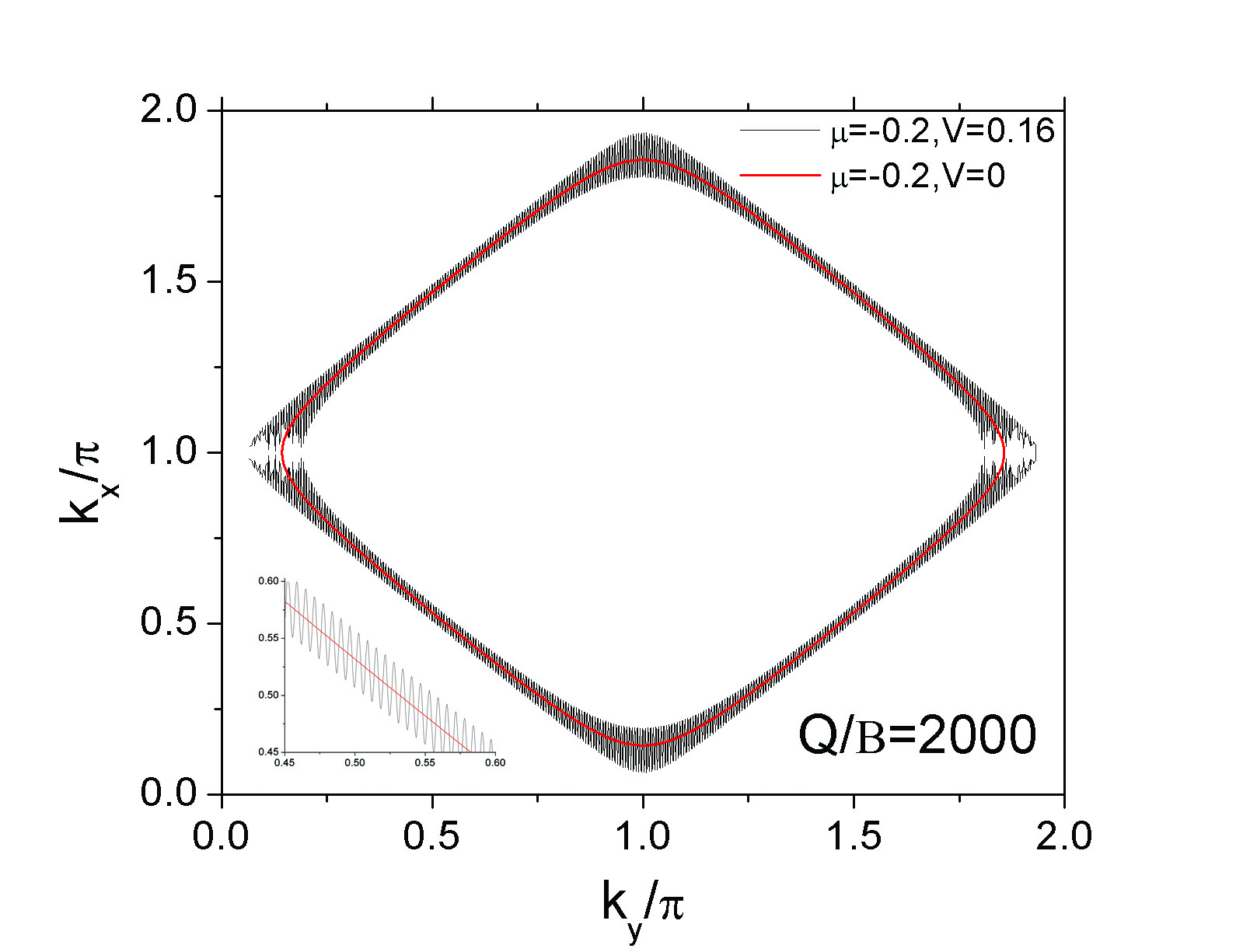}
\protect\caption{The Fermi surfaces projected to the $xy$ plane for
parameters $V=0.16$, $\mu=-0.2$ and $Q/B=80$ (upper panel) and
$Q/B=2000$ (lower panel), respectively. The inset is an enlargement
of a part of the Fermi surface. For comparison, the red curves show
the original Fermi surface without the CDW.} \label{fig15}
\end{centering}
\end{figure}

This is the regime where the three-dimensional Fermi surface is a
warped cylinder as shown in the right panel of Fig. \ref{fig9}. The
projections of the extremal three-dimensional Fermi surface cross
sections $S^{\perp}_{k}(B)$ into the $x$-$y$ plane, which we denote
as $S^{\parallel}_{k}(B)$, is related to $S^{\perp}_{k}(B)$ by
\begin{equation}
S^{\parallel}_{k}(B)= \left(\frac{B}{B^{\hskip2pt \rm eff}}\right)
S^{\perp}_k(B),
\end{equation}
and some typical projections are shown in Fig. \ref{fig15}. The
``rippled'' nature of the projected Fermi surface cross sections is
a consequence of ``neck and belly'' variations along the $\hat z$
direction induced by the $k_z$ dispersion in the effective
three-dimensional Fermi surface; despite its complexity, the total
area $S^{\parallel}_{k}(B)$ is close to that of the original
two-dimensional Fermi surface in the absence of the CDW (red curves
in Fig. \ref{fig15}), since the contributions from numerous ``neck''
and ``belly'' parts become averaged out. We therefore see that
despite the absence of truly periodic oscillations in $1/B$, there
exist approximately periodic oscillations since
$S^{\perp}_{k}(B)/\sqrt{B^2 + (Q/2\pi)^2} = S^{\parallel}_{k}(B)/B
\sim const/B$.

\begin{figure}
\begin{centering}
\includegraphics[scale=0.3]{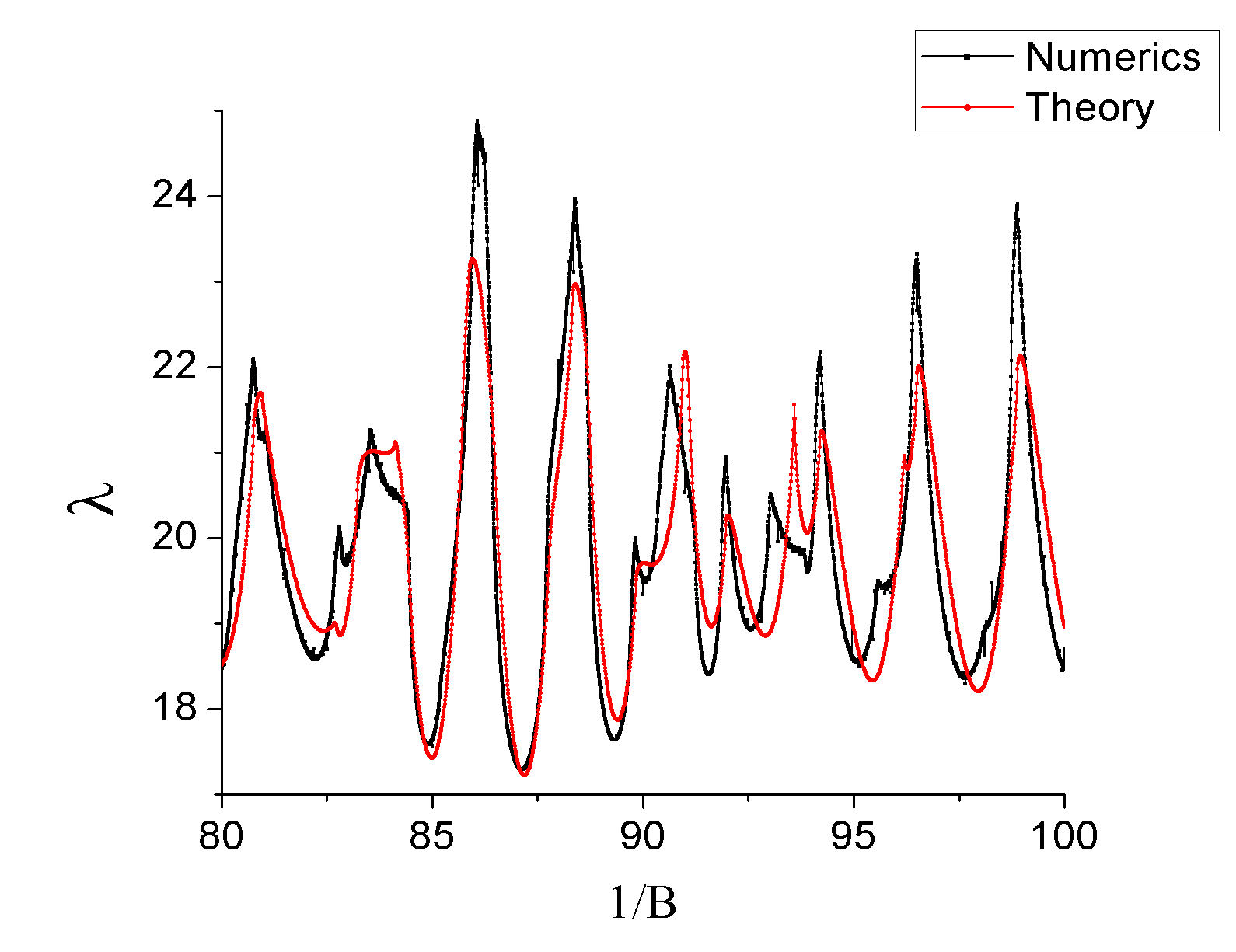}
\includegraphics[scale=0.3]{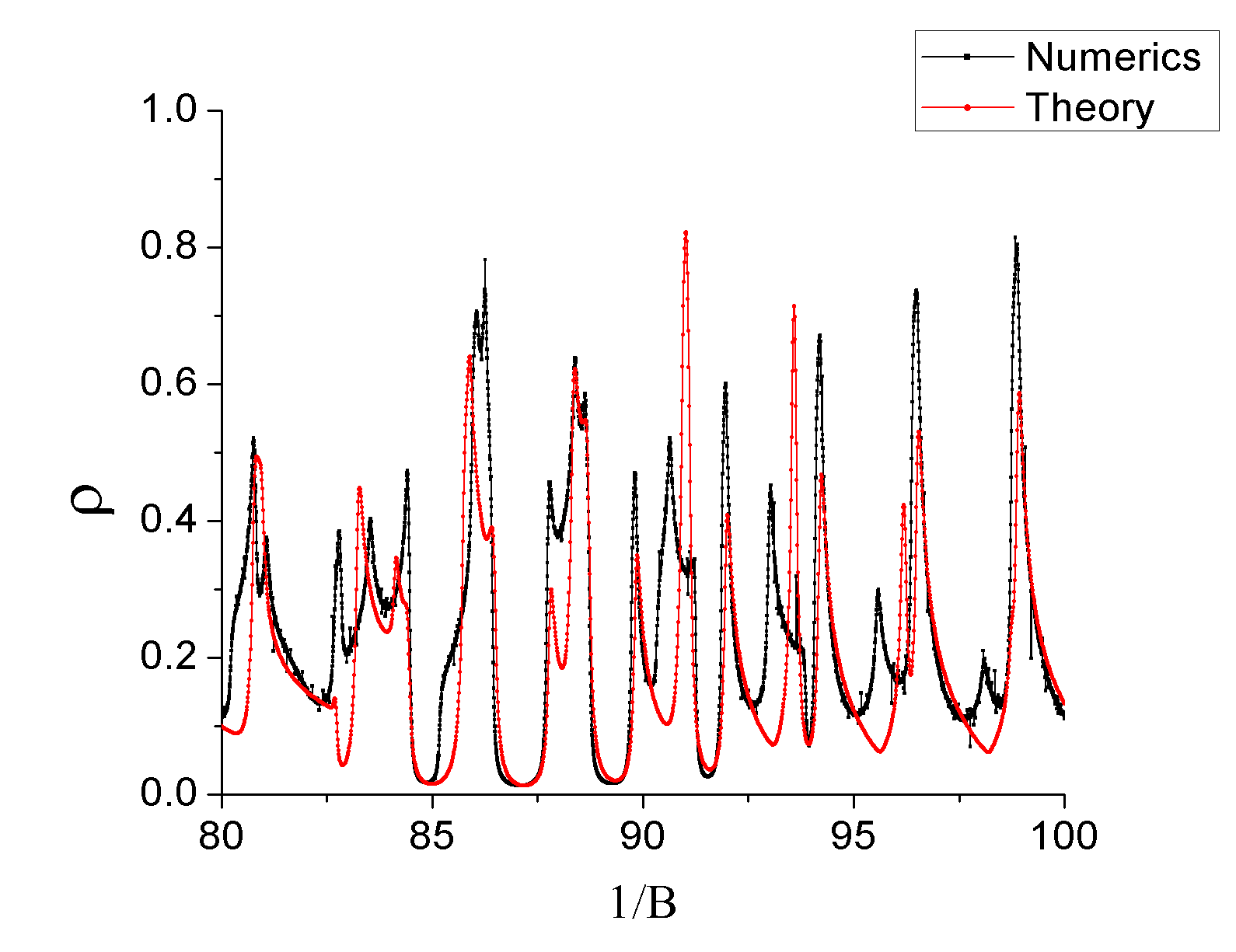}
\protect\caption{The QOs of the localization length $\lambda$ (upper
panel) and DOS $\rho$ (lower panel) as calculated numerically and
compared with theoretical simulations over a range of relatively
large magnetic field $1/B\in[80,100]$ with parameters $V=0.16$,
$\mu=-0.2$, $Q=0.4$. Additional fine structures and peaks are
observed along with the fundamental oscillation period. The system
size is $L=7.2\times 10^5$} \label{fig17}
\end{centering}
\end{figure}

Nevertheless, since there are typically two or more extremal orbits
of the three-dimensional Fermi surface, there are contributions from
multiple $S^{\parallel}_{k}(B)$ and corresponding ``fine
structures'' in the QOs. As an example, we show in Fig. \ref{fig17}
the QOs of both the localization length and the DOS with parameters
$V=0.16$, $\mu=-0.2$, $Q=0.4$ over a range of relatively large
magnetic field $1/B\in[80,100]$. While the main QO period is clearly
seen, there exist additional fine structures. As a consistency
check, the red curves in Fig. \ref{fig17} were obtained by summing
over the contributions from all cross sections according to their
respective semiclassical interference condition
$S^{\parallel}_{k}(B)/B$.

As we move to smaller values of magnetic field $B$,
$\vec{B}^{\hskip2pt \rm eff}$ becomes closer to $\hat y$ and its
perpendicular planes closer to the $xz$ plane, thus the
$S^{\perp}_{k}(B)$ cross sections span more Brillouin zones in the
$\hat z$ direction. Inevitably, the extremal
$S^{\parallel}_{k}(B)$ becomes closer and the fine structure 
 gets
suppressed. If we reduce $B$ further, however, the projected Fermi
surface becomes very complex and singular, as shown in the lower
panel of Fig. \ref{fig15}. This is where the three-dimensional
representation, while still exact, ceases to be useful due to a
catastrophic magnetic breakdown of the semiclassical approximation.

\begin{table}
\begin{tabular}{|c|c|c|}
\hline Magnetic breakdown &  Yes & No\tabularnewline \hline Original
2D system & small $Q/B$ and $V$ & large $Q/B$ and $V$\tabularnewline
\hline Dual 3D model & large $Q/B$ and $V$ & small $Q/B$ and
$V$\tabularnewline \hline
\end{tabular}
\caption{The duality between the magnetic breakdown parameter
regimes of the original two-dimensional system and its dual
three-dimensional version.}\label{table1}
\end{table}

Interestingly, there is a `duality' between the magnetic breakdown
scenarios in the original two-dimensional system and its dual
three-dimensional model: the regime of the magnetic breakdown (QO
period approximately determined by the Fermi surface without ICDW)
in the original two-dimensional system corresponds in the dual
three-dimensional model to the regime without magnetic breakdown; on
the other hand, the two-dimensional reconstructed Fermi surface
(without magnetic breakdown) is relevant when magnetic breakdown
becomes important in the dual three-dimensional system. The
qualitative results on the parameter regimes for the magnetic
breakdowns in %
the dual scenarios are summarized in Table \ref{table1}.
We discuss more details in Appendix \ref{dualmagbreak}.

\subsubsection{$V_c$ is a crossover}
While a sharp Lifshitz transition occurs in the three-dimensional
band structure, this does not imply that there is a sharp transition
in the QO spectrum at $V=V_c$. It is important to remember that for
fixed $Q$, $|\vec B^{\hskip2pt \rm eff}| \geq Q$, so magnetic
breakdown effects are significant wherever small gaps occur in the
three-dimensional band structure. Of course, this situation
necessarily arises as $V$ approaches $V_c$. Consequently, in the
exact solution, $V_c$ marks a crossover in behavior -- a crossover
that is increasingly sharp as $Q$ and $B$ get smaller.

\section{Generalizations of the duality
relations}\label{sec:generalization}

\subsection{Localization transition in an incommensurate
potential}\label{sec:localization}

The problem of a one-dimensional chain in the presence of an ICDW is
a classic problem that has been widely
studied\cite{aubry1979,sokoloff1985}. The corresponding eigenvalue
equation is known as the ``almost Mathieu equation." The dual
mapping of this problem to a two-dimensional crystal in the presence
of a magnetic field is equivalent to the Hofstadter problem with
incommensurate flux density, $B=Q/2\pi$, and with hopping matrix
elements $1$ and $V/2$ along the $\hat{x}$ and $\hat{y}$ directions,
respectively. (For $V=2$, the one-dimensional equation is known as
Harper's equation.)

\begin{figure}
\begin{centering}
\includegraphics[scale=0.30]{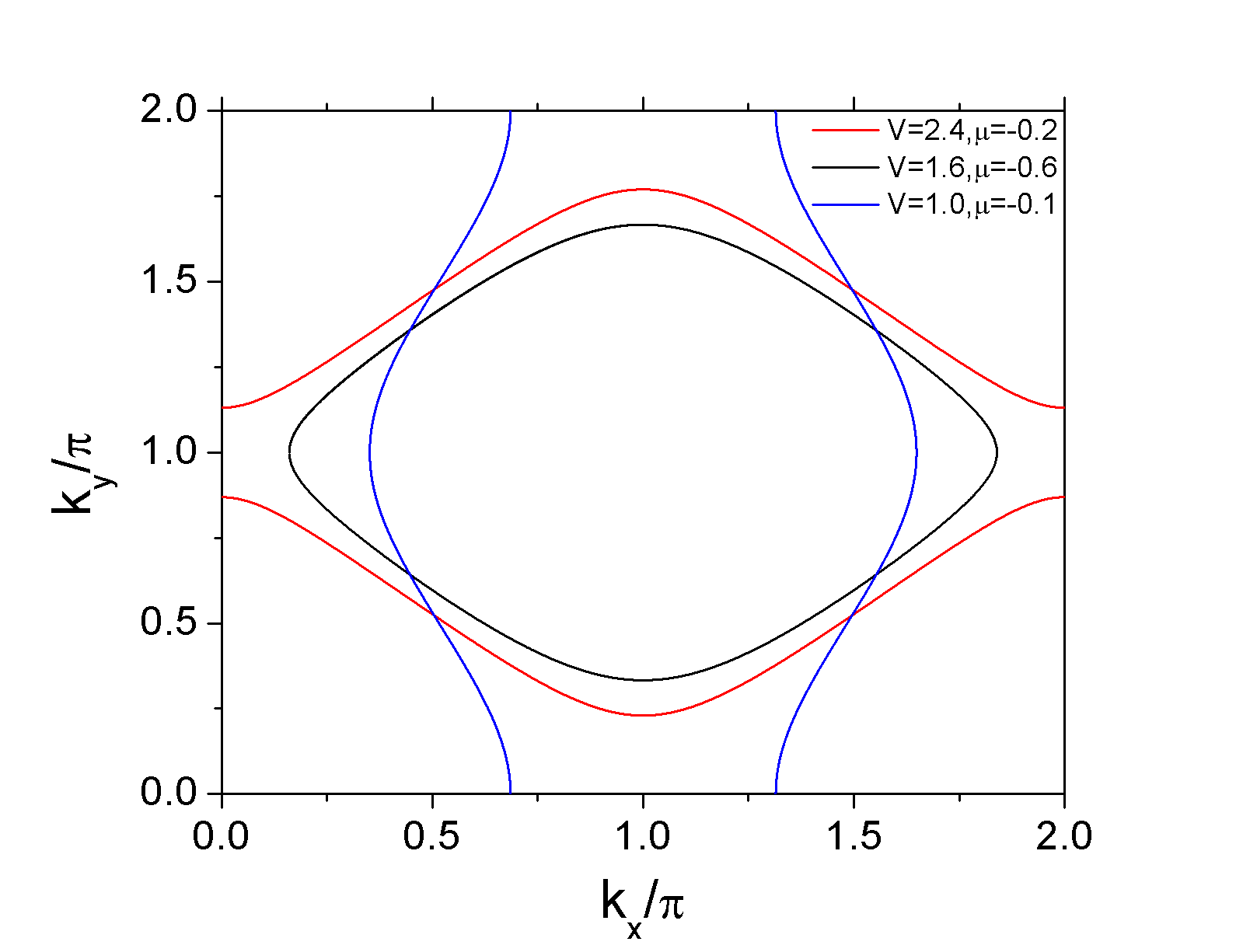}
\includegraphics[scale=0.30]{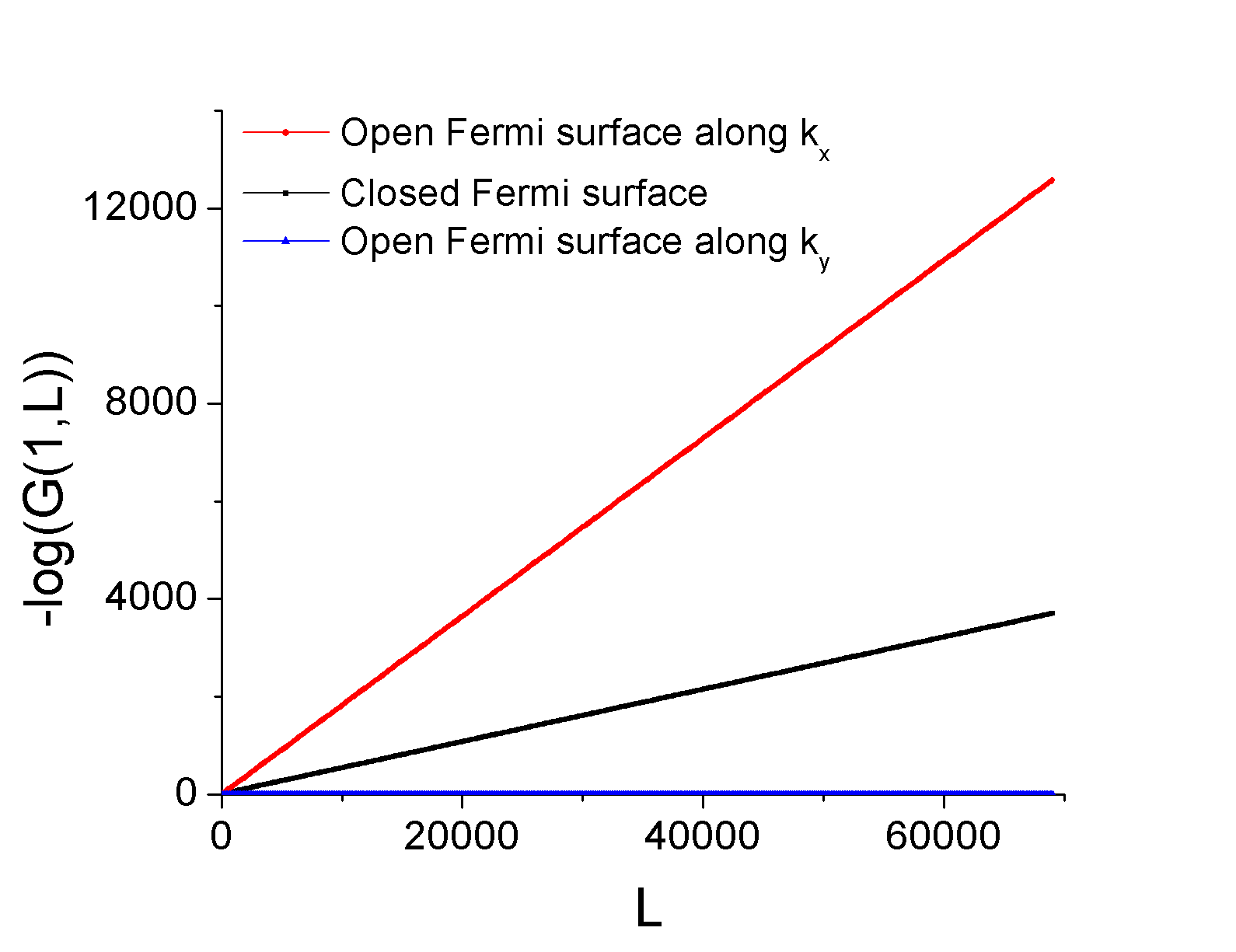}
\protect\caption{The Fermi surfaces (upper panel) and the Green's
function $-\log(G(1,L))$ versus $L$ (lower panel) with parameters
$V=1.6$, $\mu=-0.6$ (black: closed Fermi surface), $V=2.4$,
$\mu=-0.2$ (red: open Fermi surface along the $\hat x$ direction)
and $V=1$, $\mu=-0.1$ (blue: open Fermi surface along the $\hat y$
direction), respectively. An exponential suppression is observed for
the former two cases while the latter case indicates
de-localization.} \label{fig18}
\end{centering}
\end{figure}

This equation has been the subject of numerous studies focusing on
the supposed ``metal-insulator'' transition which is thought to
occur as a function of $V$; from delocalized states which occur for
small $V$ to purely localized states which occur for large $V$. In
particular, it has been claimed\cite{aubry1979,sokoloff1985} that
for $V<2$, all states are delocalized while for $V>2$ all states are
localized. We have found in our numerical solution of this problem
as well as by analyzing the semiclassical dynamics in Hofstadter
problem, that the former claim is likely incorrect - even for $V<2$
states near the band bottom are localized so long as $Q$ is truly
irrational. We discuss this aspect of the problem in detail in a
forthcoming paper,\cite{forthcoming} but here we sketch the basic
point.

In a magnetic field, the electron orbits in the two-dimensional
momentum space are on the Fermi surface, which is given by the
dispersion relation $\mu=\epsilon_{k}=2\cos k_{x}+V\cos k_{y}$.
Since the velocity of an electron at the Fermi energy is everywhere
perpendicular to the Fermi surface, the semiclassical motion is
localized in the $\hat{x}$ direction (which is the spatial direction
in the original one-dimensional problem) unless the Fermi surface is
open along the $\hat{y}$ direction. Thus, in the semiclassical
approximation, for fixed $\mu$, states at the Fermi level are
delocalized provided $V\in(-2+\left|\mu\right|,2-\left|\mu\right|)$,
and otherwise are effectively localized. To test the validity of
this semiclassical argument, we have numerically computed the
Green's function $G(1,L)$ with the same parameters used to generate
the two-dimensional Fermi surfaces in the upper panel of Fig.
\ref{fig18}, assuming an incommensurate wave vector $Q=1/31$; the
results are shown in the lower panel of Fig. \ref{fig18}. Clearly,
the semiclassical results are fully consistent with numerical
calculations, in contradiction to claims that all states are
delocalized for $V<2$ (and $\mu \neq 0$).

It is important to remember that the CDW ordering vector $Q$ plays
the role of $B^{\hskip2pt \rm eff}$ in this discussion. We have
worked with a relatively small value of $Q$ where the semiclassical
results are expected to be more or less reliable, but for larger
values of $Q$, magnetic breakdown should be exceedingly important,
especially when $V$ is smaller but close to the transition $V_c$.
Therefore, for fixed $\mu\neq 0$, the critical value of $V$ is
shifted from the semiclassical value derived above. This effect has
also been verified in our numerical calculations.

\subsection{Bidirectional ICDW in a magnetic field}

It is straightforward to generalize similar higher-dimensional dual
models and lower-dimensional numerical methods to a bidirectional
ICDW. For a $d$-dimensional hyper-cubic lattice in the presence of a
uniform magnetic field and $H=\underset{\vec k}{\sum}H_{\vec k}$
with
\begin{eqnarray}
H&=&-\underset{x,y,\vec k}{\sum} t_x a_{x+1,y,\vec k}^{\dagger}
a_{x,y,\vec k}+t_x a_{x-1,y,\vec k}^{\dagger}a_{x,y,\vec k}\\
&+&t_y e^{i\Phi_y x,\vec k}a_{x,y+1,\vec k}^{\dagger} a_{x,y,\vec
k}+t_y e^{-i\Phi_y x} a_{x,y-1,\vec k}^{\dagger}a_{x,y,\vec
k}+a_{x,y,\vec k}^{\dagger}a_{x,y,\vec
k}\nonumber\\&\times&\left[2\sum_{\nu=3}^d t_\nu\cos(\Phi_\nu x -
k_\nu)+V_x\cos\left(Q_x x+k_z\right)+V_y\cos\left(Q_y y+k_w\right)
\right]\nonumber
\end{eqnarray}
where $\vec k$ is the Bloch vector for the $d-2$ dimensions other
than the $\hat x$ and $\hat y$ directions. We have also denoted the
initial phase of the CDW along the $\hat x$ and $\hat y$ directions
as $k_z$ and $k_w$, respectively.

Similar to the duality argument in Sec. \ref{sec:model}, by summing
over both the $k_z$ and $k_w$, the above system is equivalent to a
$d+2$ dimensional effective model with two additional hoppings
$V_x/2$ and $V_y/2$ along the $\hat z$ and $\hat w$ directions, and
two additional magnetic fluxes $Q_x$ through each $xz$ plaquette,
and $Q_y$ through each $yw$ plaquette, respectively. On the other
hand, for efficient numerical calculations we can suppress the $\vec
k$ indices and consider the resulting two-dimensional system. We
leave more details to future work.

\section{Conclusions}\label{sec:conclusions}

For crystals - and this includes the case of a commensurate CDW --
the existence of a sharp Fermi surface has the profound consequence
that there exist exactly periodic oscillations in physical
quantities as a function of $1/B$ in the limit $T\to 0$, $B\to 0$,
and in the absence of disorder.  In this precise sense, the
breakdown of Bloch's theorem in the presence of an ICDW of arbitrary
magnitude $V$ eliminates the sharp definition of the Fermi surface
and the exactly periodic character of the associated QOs. For small
$V$, approximately periodic oscillations survive over parametrically
broad ranges of $B$, but inevitably as $B$ is reduced toward 0, the
period of the oscillations shifts as smaller and smaller gaps opened
in increasingly high order in perturbation theory become relevant. A
crossover in the properties occurs at $V= V_c \sim W$, beyond which
all periodic oscillations, and indeed essentially all magnetic field
dependence of the electronic structure, is eliminated.

In real materials, disorder or finite temperature $T$ limits the
range of applicability of our results. At very low fields, when
either $T$ or the Dingle temperature $T^{*}$ exceeds the cyclotron
energy $\omega_c$, all QOs are lost in a way that is independent of
the nature of the CDW order.

We would like to thank Boris Spivak, Andre Broido, Danny Bulmash,
Abhimanyu Banerjee, Yingfei Gu and Xiao-liang Qi for insightful
discussions. This work is supported by the Stanford Institute for
Theoretical Physics (YZ), the National Science Foundation through
the grant No. DMR 1265593 (SAK) and DOE Office of Basic Energy
Sciences under contract No. DEAC02-76SF00515(AM).

\newpage

\appendix

\section{Green's function and recursive relations}\label{sec:numerics}

Inspired by the recursive Green's function method, here we propose
an efficient method to calculate some targeted Green's functions and
their related physical measurable. First, it is straightforward to
derive:
\[
H_{ky}-\mu-i\delta=\left(\begin{array}{ccccc}
h_{1} & 1\\
1 & h_{2} & 1\\
 & 1 & h_{3} & 1\\
 &  & 1 & h_{4} & \ddots\\
 &  &  & \ddots & \ddots
\end{array}\right)
\]
where $h_{x}=2\cos\left(2\pi\Phi
x+k_{y}\right)+V\cos\left(Qx+\phi\right)-\mu-i\delta$. Here $\delta$
denotes a very small imaginary part to round off the singularities
and account for a finite line width for the energy levels and will
be useful later in the Green's function. For large enough systems
the detailed value of $\delta$ does not make essential changes to
our results and conclusions. In particular, the Green's function for
each $k_{y}$ is given by the matrix inverse:
\[
G_{k_{y}}=\left(\mu+i\delta-H_{k_{y}}\right)^{-1}
\]

For simpler notation, let's define:
\[
\mathcal{D}_{j}^{i}\equiv\det\left(\begin{array}{ccccc}
h_{i} & 1\\
1 & h_{i+1} & 1\\
 & 1 & h_{i+2} & \ddots\\
 &  & \ddots & \ddots & 1\\
 &  &  & 1 & h_{j}
\end{array}\right)
\]
which has the following recursive relations:
\[
\mathcal{D}_{j}^{i}=h_{j}\mathcal{D}_{j-1}^{i}-\mathcal{D}_{j-2}^{i}
\]
\[
\mathcal{D}_{j}^{i}=h_{i}\mathcal{D}_{j}^{i+1}-\mathcal{D}_{j}^{i+2}
\]
or equivalently:
\[
\mathcal{D}_{j}^{i}/\mathcal{D}_{j-1}^{i}=h_{j}-\mathcal{D}_{j-2}^{i}/\mathcal{D}_{j-1}^{i}
\]
\[
\mathcal{D}_{j}^{i}/\mathcal{D}_{j}^{i+1}=h_{i}-\mathcal{D}_{j}^{i+2}/\mathcal{D}_{j}^{i+1}
\]
with the initial conditions: $\mathcal{D}_{1}^{1}=h_{1}$,
$\mathcal{D}_{2}^{1}=h_{1}h_{2}-1$, $\mathcal{D}_{L}^{L}=h_{L}$ and
$\mathcal{D}_{L}^{L-1}=h_{L-1}h_{L}-1$.

\emph{The localization length:}

The Green's function between two ends of the system along the $\hat{x}$
direction is exponentially suppressed as the system length $L$ increases:
$\left|G\left(1,L\right)\right|\sim\exp\left(-L/\lambda\right)$.
$\lambda$ is the localization length - an important property that
characterizes transport behaviors of the system. To calculate $G\left(1,L\right)$
we need to trace over $k_{y}$:
\[
G\left(1,L\right)=\underset{y}{\sum}G\left(\left\{ 1,y\right\} ;\left\{ L,y\right\} \right)=\underset{k_{y}}{\sum}G_{k_{y}}\left(1,L\right)
\]
yet we have shown in the main text that in the presence of a small
magnetic field $\Phi$, the localization length derived from
$G\left(1,L\right)$ is equivalent to that from each individual
$G_{k_{y}}\left(1,L\right)$.

Then, by standard matrix inversion procedure, it is not hard to obtain:
\[
G_{k_{y}}\left(1,L\right)=(-1)^{L}/\mathcal{D}_{L}^{1}
\]
which can be efficiently derived using the recursive relations. We
can perform a log-linear fit with respect to $L$ for the localization
length $\lambda$.

Further, the above expression is consistent if we use
$G(1,x)\sim\exp\left(-x/\lambda\right)$ to extract $\lambda$ with a
fixed system size $L$:
$G_{k_{y}}\left(1,x\right)=(-1)^{x}\mathcal{D}_{L}^{x-1}/\mathcal{D}_{L}^{1}$.

\emph{The density of states(DOS):}

The DOS at chemical potential $\mu$ is another important physical
quantity that we mainly focus on:
\[
\rho\left(\mu\right)=-\frac{1}{\pi LL_{y}}\underset{x,k_{y}}{\sum}\mbox{Im}G_{k_{y}}\left(x,x\right)=-\frac{1}{\pi L}\underset{x}{\sum}\mbox{Im}G_{k_{y}}\left(x,x\right)
\]

Again, we have shown in the main text that the DOS averaged over
$k_{y}$ is equivalent to that for each specific $k_{y}$. We also
note that:
\begin{eqnarray}
&
&-G_{k_{y}}\left(x,x\right)=\mathcal{D}_{x-1}^{1}\mathcal{D}_{L}^{x+1}/\mathcal{D}_{L}^{1}\nonumber\\&=&\mathcal{D}_{x-1}^{1}\mathcal{D}_{L}^{x+1}/\left(\mathcal{D}_{x-1}^{1}\mathcal{D}_{L}^{x+1}h_{x}-\mathcal{D}_{x-2}^{1}\mathcal{D}_{L}^{x+1}-\mathcal{D}_{x-1}^{1}\mathcal{D}_{L}^{x+2}+2\right)\nonumber\\&\eqsim&\left(h_{x}-\mathcal{D}_{x-2}^{1}/\mathcal{D}_{x-1}^{1}-\mathcal{D}_{L}^{x+2}/\mathcal{D}_{L}^{x+1}\right)^{-1}
\end{eqnarray}
where in the last step we have neglected the term of $2$ since
$2/\mathcal{D}_{L}^{1}$ is exponentially small.
$-G_{k_{y}}\left(x,x\right)$ may be efficiently extracted with the
recursive relations. Both $\lambda$ and $\rho$ should oscillate as a
function of $1/B$ with periods given by the cross-section area of
the Fermi surface.

\section{Duality in magnetic breakdown scenarios}\label{dualmagbreak}

As we have shown in the main text, for $V\ll W$ and $B\ll Q$, the
dual three-dimensional Fermi surface cross section's projected area
onto the $xy$ plane, $S_{k}^{\parallel}$, is almost invariant and
asymptotically close to the original Fermi surface, therefore we
expect QOs with a period $\Delta(1/B)$ in the regime where the
electron semiclassical orbit restores the original Fermi surface.
The magnetic breakdown in the original two-dimensional system are
given by the dual three-dimensional model with no magnetic
breakdown.

On the other hand, in the limit $B\rightarrow 0$, the magnetic
breakdown in the original two-dimensional system gets washed out and
the QOs are dominated by the CDW reconstructed Fermi surface.
However, in this limit, the Fermi surface in the dual
three-dimensional model also gets increasingly complex and singular
(see Fig. \ref{fig15} lower panel), and magnetic breakdown becomes
essential. The electrons can tunnel between the ripples given
$\ell\delta k>1$ where $\ell$ is the magnetic length (that is almost
constant since $B_{\hskip2pt \rm eff}\sim Q/2\pi$) and $\delta k$ is
the distance between two neighboring ripples, thus denser (larger
$Q/B$) and stronger (larger $V$) ripple structures are favorable to
the magnetic breakdown, which is exactly opposite to the regimes of
the original two-dimensional system, see Table \ref{table1}.

\begin{figure}
\begin{centering}
\includegraphics[scale=0.15]{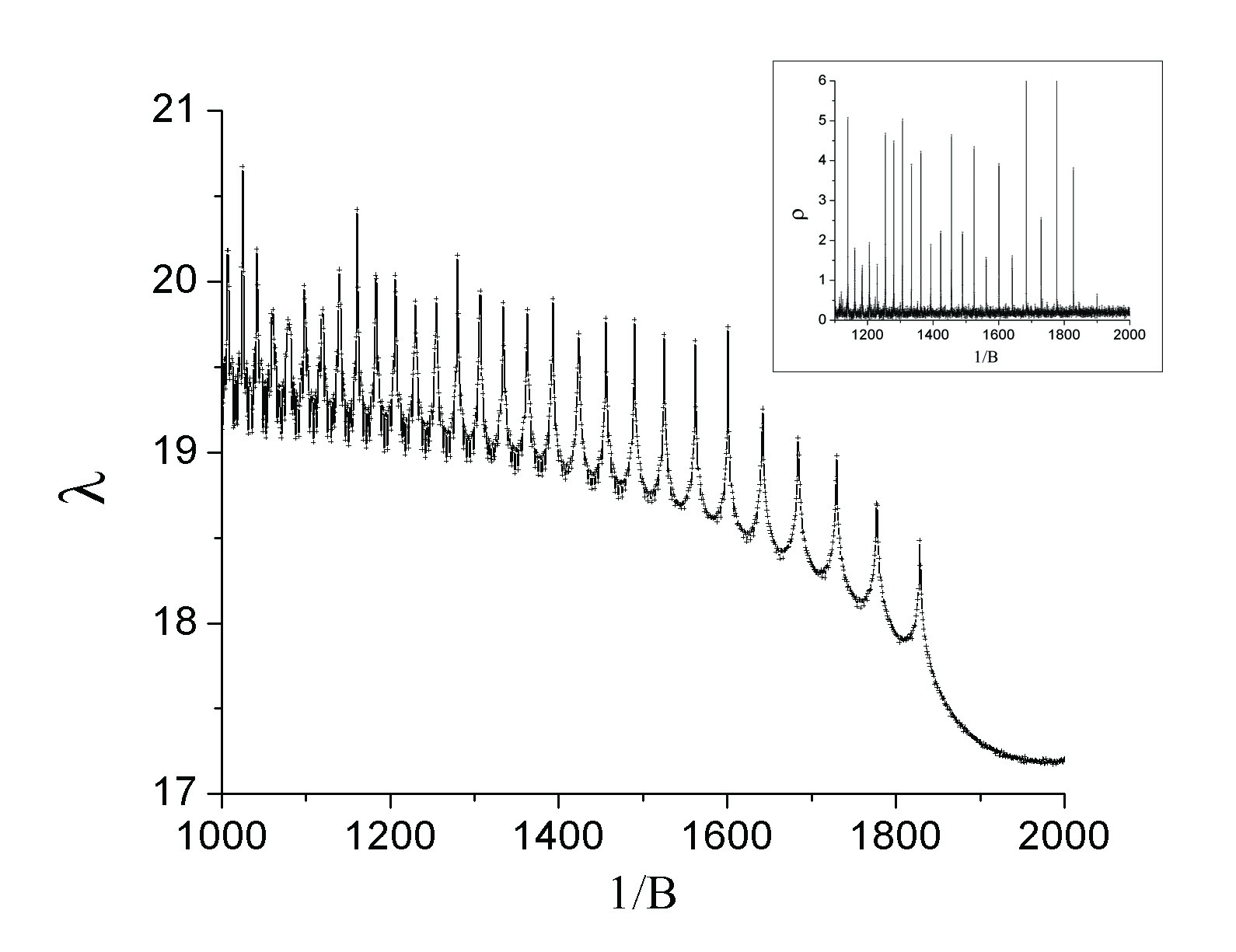}
\includegraphics[scale=0.15]{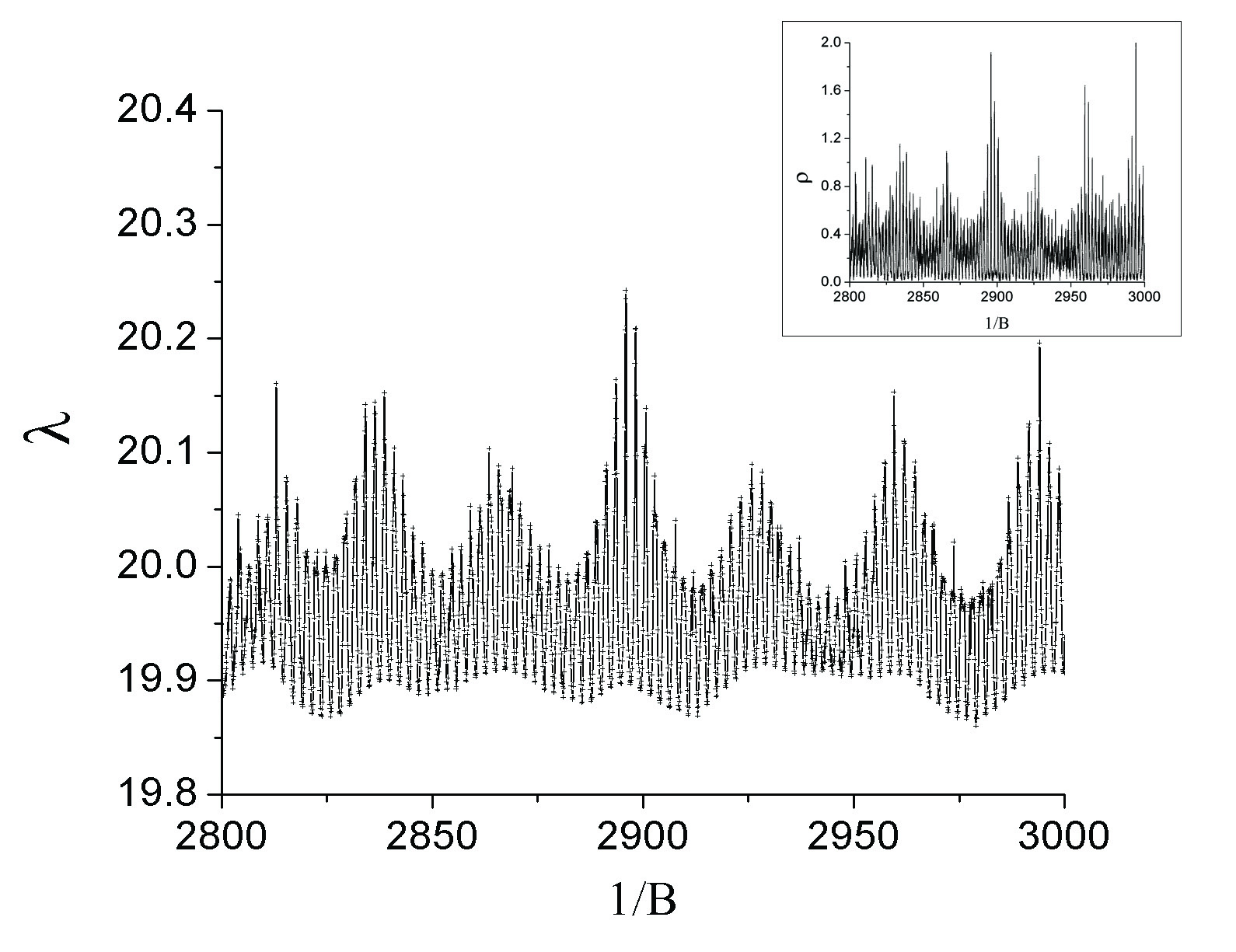}
\protect\caption{Upper panel: The localization length $\lambda$ and
the DOS $\rho$ (insets) with parameters $V=0.16$, $\mu=-0.2$ and the
ratio $Q/B=2000$ fixed over a range of $1/B\in[1000,2000]$
($Q\in[1,2])$. The oscillation becomes aperiodic and vanishes as $B$
gets smaller. Lower panel: if we reduce $B$ even further:
$1/B\in[2800,3000]$, the periodicity of the QO is well restored. The
system size is $L=7.2\times 10^5$.} \label{fig16}
\end{centering}
\end{figure}

To manifest the impact of magnetic breakdown in the dual
three-dimensional model, we present in Fig. \ref{fig16} the
numerical results on both the localization length and the DOS with a
fixed ratio $Q/B=2000$ for the Fermi surface in Fig. \ref{fig15}
lower panel. The resulting behaviors in the range
$1/B\in[1000,2000]$ are no longer periodic and essentially different
from those in Fig. \ref{fig11}. This can be explained as the
following: the magnetic breakdown introduces tunnelings between the
ripples and effectively increases the area enclosed by the electron
trajectory $S_k^{\parallel}(B)$. As we lower the magnetic field
$B_{\hskip2pt \rm eff}\propto B$, the distance the electrons can
tunnel through is reduced, thus $S_k^{\parallel}(B)$ is a decreasing
function of $1/B$. When this happens, the orbital phase factor
$S_k^{\parallel}(B)/B$ is slower than $1/B$, the periodicity in
$1/B$ is lost and the interval between the oscillations becomes and
larger, see Fig. \ref{fig16} upper panel. On the other hand, when
the magnetic field $B_{\hskip2pt \rm eff}\propto B$ is reduced
further so that the magnetic breakdown becomes unimportant in the
dual three-dimensional model, the periodicity should be restored.
This is exactly what we observe in Fig. \ref{fig16} lower panel.

\end{document}